\documentclass{article}
\usepackage{amsmath}
\usepackage{subcaption}
\usepackage{titling}
\usepackage{graphicx}
\usepackage[table, xcdraw]{xcolor}
\usepackage{svg}
\usepackage[utf8]{inputenc}
\usepackage{multicol}
\usepackage[top=25mm,bottom=30mm,inner=20mm,outer=35mm]{geometry}
\usepackage{mathptmx}
\usepackage[linecolor=red,backgroundcolor=red!25,bordercolor=red,textwidth=30mm]{todonotes}
\usepackage{blindtext}
\usepackage[font=small, labelfont=bf]{caption}
\usepackage{authblk}
\usepackage{xr}
\usepackage[section]{placeins}
\usepackage[colorlinks,linkcolor=red,anchorcolor=black,citecolor=green]{hyperref}
\usepackage{abstract}
\usepackage{cite}

\usepackage{soul}
\usepackage{boldline}
\definecolor{lightgray}{gray}{0.9}
\usepackage{algorithm}
\usepackage[noend]{algpseudocode}
\usepackage{cleveref}
\usepackage{comment}
\usepackage{makecell}
\usepackage{harpoon}
\usepackage{multirow}
\usepackage{amsfonts}

\let\nvec\vec
\def\vec#1{\nvec{\vphantom t\smash{#1}}}

\title{Image prediction of disease progression by style-based \textcolor{blue}{manifold} extrapolation}

\author[1,*]{Tianyu Han}
\author[2,3,4]{Jakob Nikolas Kather}
\author[5]{Federico Pedersoli}
\author[5]{Markus Zimmermann}
\author[5]{Sebastian Keil}
\author[5]{Maximilian Schulze-Hagen}
\author[5]{Marc Terwoelbeck}
\author[5]{Peter Isfort}
\author[6]{Christoph Haarburger}
\author[7,8,9]{Fabian Kiessling}
\author[1,7,8,*]{Volkmar Schulz}
\author[5]{Christiane Kuhl}
\author[5 $\dagger$]{Sven Nebelung}
\author[5,*, $\dagger$]{Daniel Truhn}
\date{}
\affil[1]{\small Physics of Molecular Imaging Systems, Experimental Molecular Imaging, RWTH Aachen University, Germany}
\affil[2]{\small Department of Medicine III, University Hospital RWTH Aachen, Aachen, Germany}
\affil[3]{\small Pathology and Data Analytics, Leeds Institute of Medical Research at St James’s, University of Leeds, Leeds, UK}
\affil[4]{\small Medical Oncology, National Center for Tumor Diseases, University Hospital Heidelberg, Heidelberg, Germany}
\affil[5]{\small Department of Diagnostic and Interventional Radiology, University Hospital Aachen, Germany}
\affil[6]{\small CheckupPoint GmbH, Munich, Germany}
\affil[7]{\small Comprehensive Diagnostic Center Aachen (CDCA), University Hospital RWTH Aachen, Aachen, Germany}
\affil[8]{\small Fraunhofer Institute for Digital Medicine MEVIS, Bremen, Germany}
\affil[9]{\small The Institute for Experimental Molecular Imaging, RWTH Aachen University, Germany}
\affil[$\dagger$]{\small Both authors contribute equally}
\affil[*]{Correspondence should be addressed to V.S (vschulz@ukaachen.de), D.T (dtruhn@ukaachen.de), and T.H (tianyu.han@pmi.rwth-aachen.de)}

\begin{document}
\maketitle
\begin{abstract}
\textbf{\textcolor{blue}{
	Disease-modifying management aims to prevent deterioration and progression of the disease, not just relieve symptoms. 
	Unfortunately, the development of necessary therapies is often hampered by the failure to recognize the presymptomatic disease and limited understanding of disease development.
	We present a generic solution for this problem by a methodology that allows the prediction of progression risk and morphology in individuals using a latent extrapolation optimization approach. 
	To this end, we combined a regularized generative adversarial network (GAN) and a latent nearest neighbor algorithm for joint optimization to generate plausible images of future time points.
	We evaluated our method on osteoarthritis (OA) data from a multi-center longitudinal study (the Osteoarthritis Initiative, OAI). 
	With presymptomatic baseline data, our model is generative and significantly outperforms the end-to-end learning model in discriminating the progressive cohort. 
	Two experiments were performed with seven experienced radiologists. 
	When no synthetic follow-up radiographs were provided, our model performed better than all seven radiologists. 
	In cases where the synthetic follow-ups generated by our model were available, the specificity and sensitivity of all readers in discriminating progressors increased from 72.3\% to 88.6\% and from 42.1\% to 51.6\%, respectively. 
	Our results open up a new possibility of using model-based morphology and risk prediction to make predictions about
	future disease occurrence, as demonstrated in the example of OA.
}}
\end{abstract}

\section*{Introduction}
\begin{figure}[h!]
	\centering
	\scalebox{1.0}{
		\includegraphics[trim=40 530 390 0, clip, width=\textwidth]
		{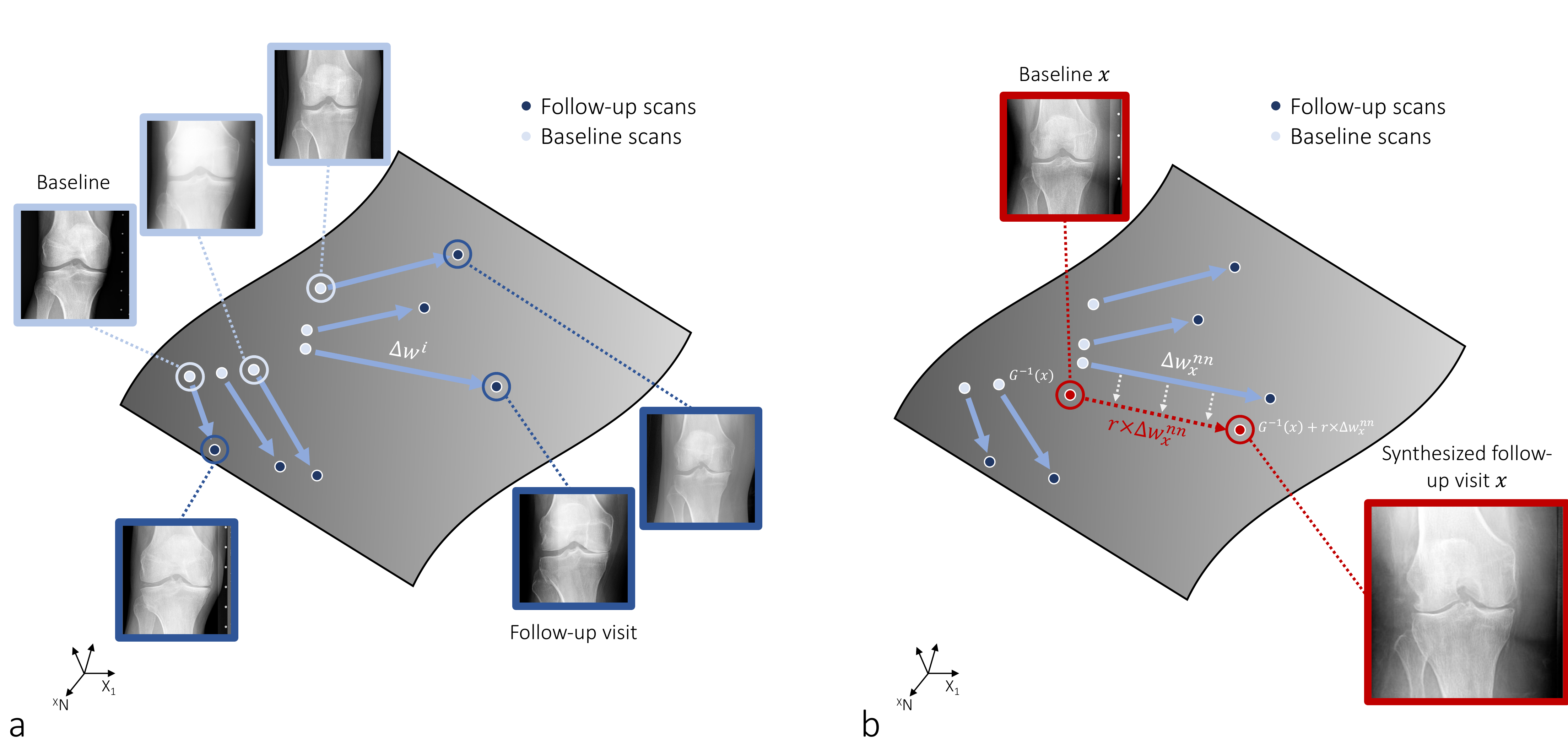}}
	\caption{\label{fig:overview}\textbf{Synthesizing follow-up radiological scans via latent nearest neighbor extrapolation.}
		(a) Encode possible OA progression scenarios from individual participants to latent trajectories in the learned manifold. 
		Knee radiographs (with a resolution of 256$\times$256) are mapped to 512-dimensional vectors in the latent space $w$. 
		OA onset or progression is represented by computing the vector $\Delta w$ that connects the baseline (light blue dots) to the follow-up radiograph (dark blue dots) of each patient. 
		Longitudinal imaging data from OAI dataset populates our latent space. 
		(b) Generation of the predicted follow-up radiographs. 
		When predicting the future image morphology of a baseline knee radiograph $x$. 
		It is primarily mapped to the 512-dimensional latent space ($\text{G}^{-1}(x)$).
		Its baseline nearest neighbors (nn) are identified by calculating normalized cosine distances, and the vector mapping the prior to the follow-up image is determined via equation \ref{equ:vectorfield}. 
		The future appearance of the knee $x$ is then generated by feeding $\text{G}^{-1}(x)+\Delta w_x^{\text{nn}}$ to the StyleGAN G, where $\Delta w_x^{\text{nn}}$ is the latent vector difference between the prior and the follow-up scans of its nn.}
\end{figure}

End-to-end learning achieved enormous success in a large number of challenges in autonomous driving \cite{grigorescu2020survey}, image recognition \cite{krizhevsky2012imagenet}, and even unfolding complicated proteins \cite{senior2020improved, jumper2021highly}. 
Similarly, in disease detection and diagnosis, the application of deep learning models in combination with medical images has also shown impressive results and might facilitate personalized diagnosis, decision management, and therapy planning \cite{lundberg2018explainable}. 
Unlike in traditional computer vision (CV) datasets such as CIFAR \cite{recht2018cifar} and ImageNet \cite{deng2009imagenet}, there are vast amounts of longitudinally acquired medical image data because of prior examinations that contain valuable implicit image information about the onset or progression of a particular disease \cite{petersen2010alzheimer, menze2014multimodal, casey2018adolescent, eckstein2012recent}.
Such longitudinal datasets perfectly fit the task of understanding and preventing the development of neurological disease \cite{petersen2010alzheimer}, macular degeneration \cite{yim2020predicting}, and OA \cite{eckstein2012recent}.
As previously reported, the understanding of the pathogenesis in OA is limited: it remains an enigmatic condition and so far treatment mainly consists of pain management with a joint replacement for end-stage disease \cite{bijlsma2011osteoarthritis}.
It's essential for the management of OA that early detection and prevention at a reversible stage can be achieved \cite{Martel-Pelletier2016}. 
Patients can be treated accordingly, either surgically or non-surgically, to at least halt disease progression.
Moreover, early disease stages may be susceptible to new therapeutic approaches \cite{conaghan2013osteoarthritis, arden2006osteoarthritis}.
However, with current standard clinical tools differentiation remains difficult between patients who require preventive measures - that is those with a high probability of progression - and those who do not have to be treated. 
Predicting the future state of a progressive disease using end-to-end learning is challenging as
scans of the same patient at different time point will be considered jointly and therefore lead to a significant reduction of the dataset size,
and the common missing labels in the longitudinal data lead to discontinuous and heterogeneous patient trajectories. 
In this work, we utilize a regularized GAN \cite{goodfellow2014generative, karras2019style, karras2020analyzing} to learn the latent temporal trajectories of patients whose longitudinal knee joint radiographs were used to infer model-based prognostication of OA onset and progression.
\textcolor{blue}{
The training of our system as shown in Fig. \ref{fig:system} can be performed in a self-supervised fashion, i.e., without the need for any diagnostic labeling.
}
Specifically, the relevant information in radiographs is represented in a low-dimensional manifold (gray surface in Fig. \ref{fig:overview}) learned by the regularized generator, patient trajectory in this space over time can be tracked feasibly.
We found that synthesized follow-up radiographs can boost the sensitivity and specificity of radiologists' prognoses about the progressive cohort solely based on presymptomatic baselines.
Our concept proposes how model-assisted prediction of OA progression risk and possible synthesized follow-up data can facilitate presymptomatic detection and disease-modifying management in the future.

\section*{Results}
\subsection*{Prediction of OA progression from presymptomatic baselines} \label{sec:pred}
As shown in Fig. \ref{fig:overview}a, we first back map training knee radiographs to the low-dimensional manifold by inverting the generator $\text{G}^{-1}(x)$.
The model's prediction is solely based on evolution trajectories of known data points, i.e., participants.  
To construct such a latent trajectory of individual participants, we associate every mapped radiograph (blue points in Fig. \ref{fig:overview}a) with its accompanied acquisition time-stamp and construct the vector field that connects baseline and follow-up visits (blue arrows in Fig. \ref{fig:overview}a).
In the prediction phase (Fig. \ref{fig:overview}b),  we compute the latent vector of the knee of interest $x$ and further find its nearest neighbors within all other latent embeddings using the normalized cosine distance (equation \ref{equ:distance}).
The predicted knee image can be generated by feeding the extrapolated vector $\text{G}^{-1}(x)+\Delta w_x$ to the trained G, where the extrapolation $\Delta w_x$ is computed based on the empirical progression of the latent vector of the neighboring knees (equation \ref{equ:vectorfield}).
We tested our approach over a large time window of eight years on 1,290 patients from the OAI dataset \cite{eckstein2012recent} that had not been seen by the model before. 
Exemplary real baseline radiographs which exhibit no visual sign of OA, their real follow-up radiographs, and their corresponding synthesized counterparts (generated based only on the real baseline) are shown in Fig. \ref{fig:manifold}a and b.
For patients exhibiting OA progression (Fig. \ref{fig:manifold}a), the generated synthesized images demonstrate classical signs of OA, most notably joint space narrowing (as indicated by the red frames and arrows), which is not present in non-progressing knee joints (Fig. \ref{fig:manifold}b). 

\begin{figure}[h!]
	\centering
	\scalebox{1}{
		\includegraphics[trim=50 220 60 40, clip, width=\textwidth]
		{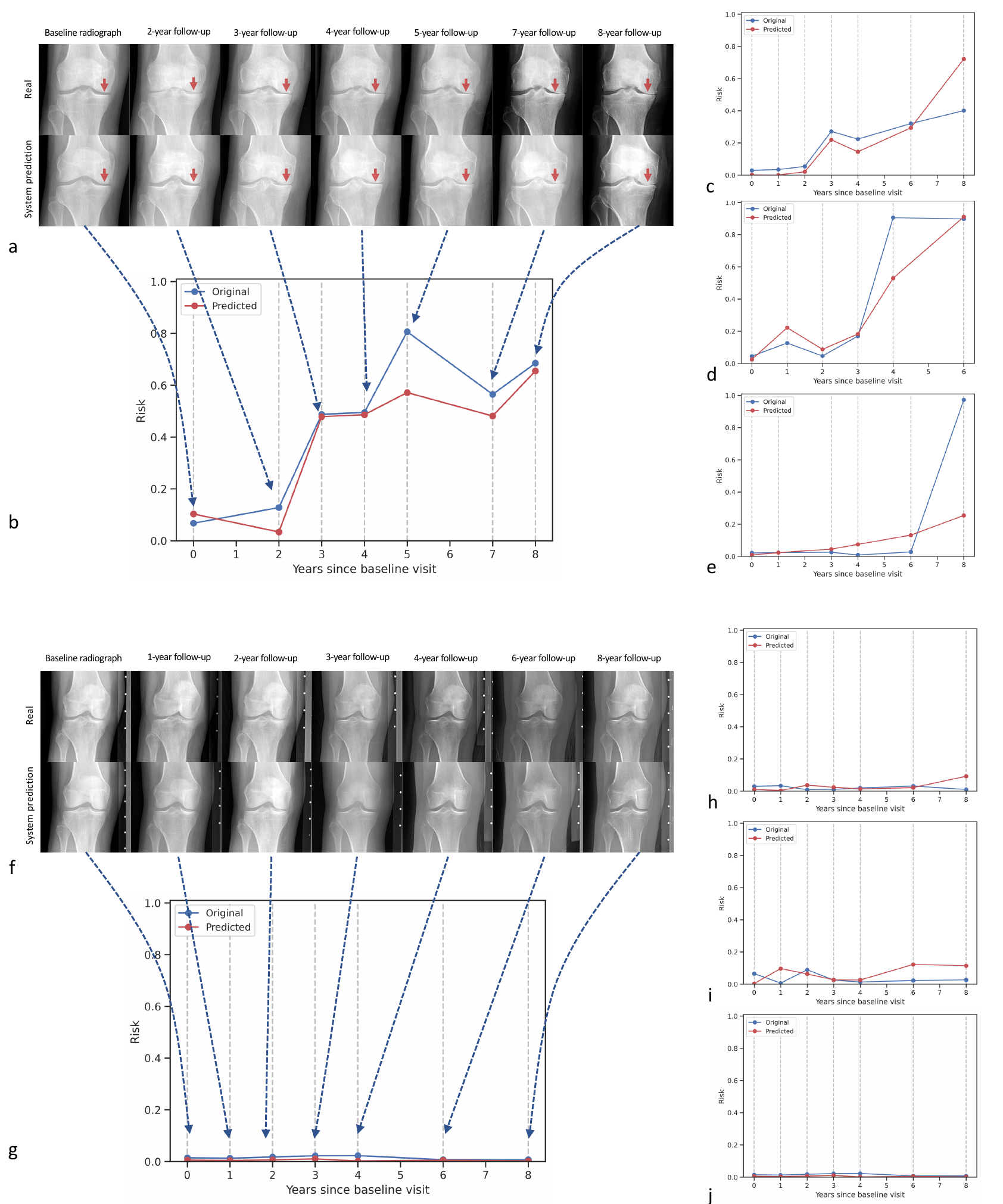}}
	\caption{\label{fig:vis_both}\textbf{Development of OA and non-OA knees exhibit distinct trajectories and are correctly predicted by our system.}
		In multiple follow-up exams (a) and (b), rapid progression of OA on the left knee of a 64-year-old male participant was observed and correctly predicted by our model.  
		The risk of OA progression in (b) is computed by comparing the baseline and subsequent follow-up radiographs as shown in Fig. \ref{fig:roc}b. 
		Such a risk trajectory clearly distinguishes progressors and normal participants as shown in (a-e) and (f-j). 
		Importantly, our model can correctly predict long-range (up to 8-years) intra-articular changes, namely narrowing of the joint space and subchondral sclerosis (indicated by red arrows in (a)), when compared with the true multi-point follow-up scans.  
		Examples in (c) and (d) further demonstrate the high relevance of the ground truth and predicted progression trajectory. 
		In (e), we also observed that the model is able to correctly predict a progression but slower than the true development that the participant suffers. 
		In contrast, when inferencing normal knees from (f) to (j), our prediction shows no sign of OA throughout 8-years follow-up studies and aligned well with ground-truth radiographs. 
	}
\end{figure}

\textcolor{blue}{
	Our model is capable of predicting the image appearance of a knee radiograph several years into the future (Fig. \ref{fig:vis_both} and \ref{fig:vis}).
	To visualize the future OA onset or progression, we generate synthesized radiographs on all follow-up time points (based solely on the baseline radiograph).
	We observe, from Fig. \ref{fig:vis_both}a and f, progressive and normal knees show distinct trajectories. 
	Monotonic increases in risk scores indicating positive OA progressions are observed consistently both in real and predicted radiographs when the testing participant comes from the progressive cohort (Fig. \ref{fig:vis_both}a to e). 
	In Fig. \ref{fig:vis_both}e, we also find the model is able to correctly predict a progression but it's slower than the true development that the participant suffers.
	Interestingly, we also notice that the synthesized follow-up radiographs (see predicted final radiograph in Fig. \ref{fig:vis}a and b) - in accordance with the real radiographs - visualize an increase in soft tissues around the knee as a sign of weight gain, often a cause for OA onset or progression due to increased mechanical load.
	In contrast, normal participants with no sign of OA maintain low and stable risk values throughout the entire study as shown in Fig. \ref{fig:vis_both}f to j.
}

\subsection*{OA progression risk quantification} \label{sec:risk}
\begin{figure}[h!]
	\centering
	\scalebox{1.0}{
		\includegraphics[trim=0 200 350 0, clip, width=\textwidth]
		{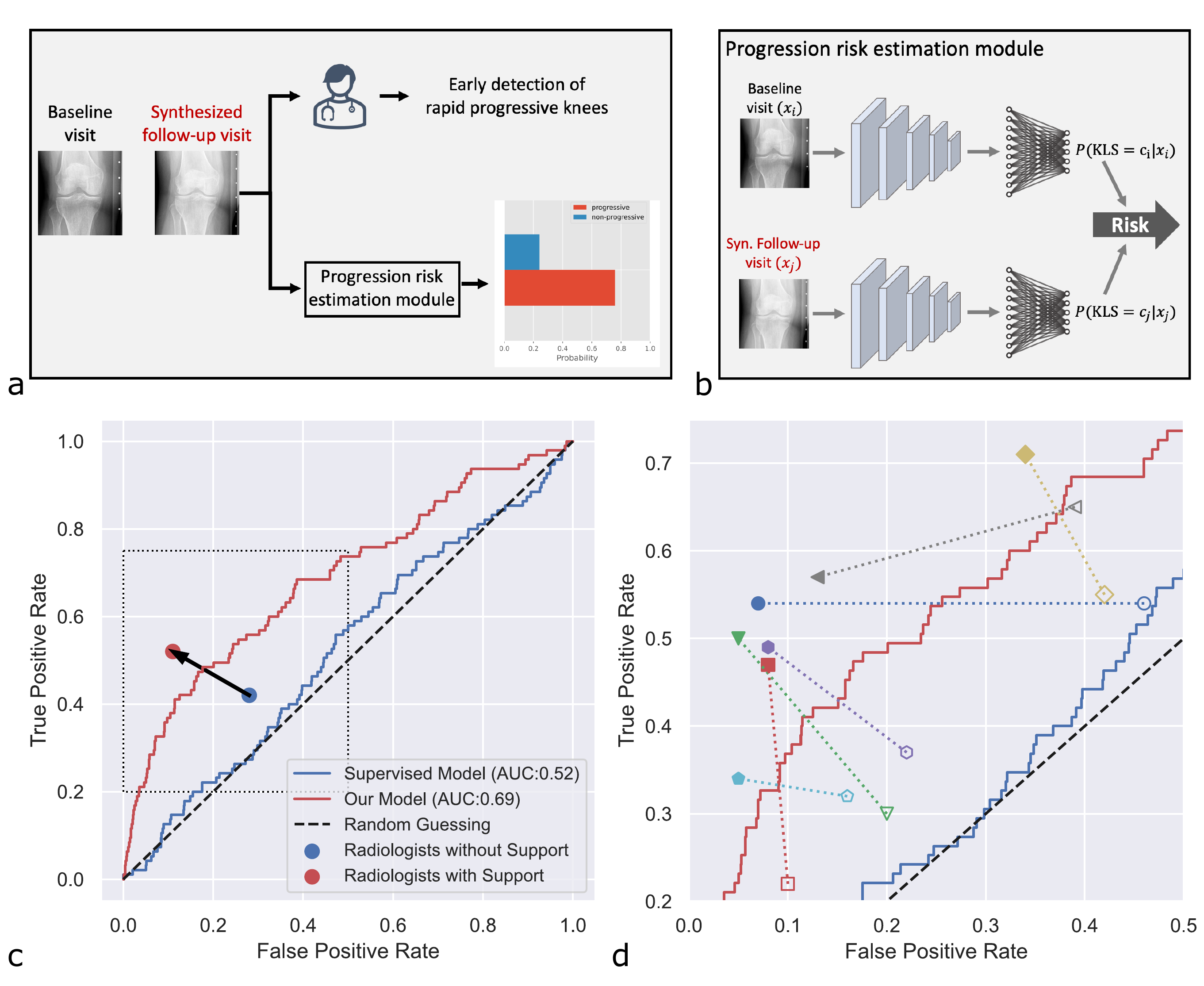}}
	\caption{\label{fig:roc}\textbf{Performance of radiologists and models in distinguishing OA progressive cohort.}
		(a) A schematic workflow of the progression risk is assessed by the trained classifier and by radiologists.
		(b) Details of our progression risk estimation.
		In this module, a pair of scans consisting of the baseline and a follow-up visit is evaluated by a KLS classifier.
		Given $p(\text{KLS}=c_i|x_i)$ and $p(\text{KLS}=c_j|x_j)$, the progression risk $p(y=1|x_i, x_j)$ is computed via equation \ref{equ:joint}.
		(c) ROC curves of the supervised classification network (blue) and our approach (red). 
		The proposed method outperforms the supervised model in terms of ROC-AUC. 
		The average performance of all radiologists without and with the support of the model is plotted as a blue and red dot, individually. 
		Both sensitivity and specificity are increased when synthesized follow-up scans by our model are provided (black arrow).
		(d) Zoomed-in region of the dotted rectangular area of the ROC (as outlined in c) with individual radiologists represented by open shapes (without model support) and filled shapes (with model support). 
		The performance of our model is superior to that of clinical experts given a mid-resolution baseline radiograph. 
		However, the prognostic performance of radiologists can also be greatly boosted by integrating our system into the loop, as shown in dashed connection lines in (d). 
		In addition, we observe that the varying performance among individual radiologists tends to reach a consistent level when synthesized follow-up scans are provided.
	}
\end{figure}

The task of estimating progression risks solely based on features provided by the baseline radiographs is challenging, even for experienced radiologists.
Progressors were participants for whom Kellgren-Lawrence score (KLS) increased more than one grade between prior and follow-up visits. 
To compare what is achievable with supervised learning based on the existing dataset, we finetuned a ResNet-50 classifier pre-trained on ImageNet dataset \cite{deng2009imagenet} that tries to identify progressors based on baseline radiographs in a supervised end-to-end manner. 
\textcolor{blue}{
In Equation \ref{equ:joint}, we obtain the progression risk by using a classifier that outputs the probability of KLS from input radiographs (see Fig \ref{fig:roc}a and b).  
Using the training scheme proposed in \cite{han2021advancing, tsipras2018robustness}, we constrained the classifier to make a prediction on the OA stage based on disease-relevant features. 
In Fig \ref{fig:grads}c and d, the comparison of gradients with respect to input pixels generated from both the used and a standard classifier verifies the above assumption.
}

We found that the classifier is no better than random guessing in predicting OA onset or progression, i.e., the area under the receiver operating characteristic curve (ROC-AUC) was 0.52.
This is partly due to the limited data and due to the difficulty of the task. 
However, our model achieved a substantially higher ROC-AUC of 0.69 on the OAI test set (Fig. \ref{fig:roc}c). 
A more detailed summary of performance metrics and confidence intervals (CI) can be found in Supplementary Table \ref{Table: ourvssup}. 
To test if our algorithm generalizes well on external data, we evaluated the predictive model's performance (as trained using the OAI data) on the independent Multi-center Osteoarthritis Study (MOST, n=3,015 participants), label distribution listed in Table \ref{Table: xrays_meta}) using our method in Fig. \ref{fig:most}.
As shown in Table \ref{Table: ourvssupmost} and in consistence with the data on the OAI dataset, our method achieves a significantly higher ROC-AUC of 0.64 (p-value $<$ 0.001, determined using bootstrapping) than its supervised counterpart with a ROC-AUC of 0.54.

\subsection*{Model assisted progressive cohort identification}

Specific treatments/drugs aim for preventing OA progression are currently not widely available due to assessing future OA evolution based on imaging data is challenging and therefore not routinely performed by radiologists.
To validate the effectiveness of our system in detecting the progressive cohort at an early stage, we designed two reader tests:
A. predict whether the patient will have OA progression based on the present radiograph only.
B. predict whether the patient will have OA progression based on the present radiograph and the synthesized future radiograph generated by our model 
\textcolor{blue}{(shown in Fig. \ref{fig:roc}a).}
In total, seven experienced clinical radiologists (with 4, 7, 6, 8, 17, 8, and 8 years of experience in musculoskeletal radiology) were asked to participate in both experiments.
In experiment A, most radiologists were not able to reliably differentiate between the two groups, except for the radiologist with 17 years of experience, see Fig. \ref{fig:roc}d and Fig. \ref{fig:readerconfusion}.
We found that the performance among radiologists varied significantly: sensitivity ranged from 22\% to 65\% and specificity ranged from 54\% to 90\% (see Fig. \ref{fig:roc}d and Table \ref{Table: readers}).
This is not surprising given the fact that radiologists have very limited experience with predicting the course of OA over many years.
In test B, results were substantially improved by the additional informative presentation of the synthesized future radiograph.
As shown in Fig. \ref{fig:roc}d and Table \ref{Table: readers}, both sensitivity and specificity consistently improved to ranges of 34\% to 71\% and 66\% to 95\% and all radiologists reached a significantly better performance as quantified by the ROC-AUC.
Comparing these results to the fully automated assessment with our model, we find that our algorithm outperforms almost all radiologists in test A: only the most senior radiologist is on par with our model. 
If supported by the model, radiologists -both individually and overall- consistently outperform the model (Test B, Fig. \ref{fig:roc}d), which highlights the fact that human and artificial intelligence are complementary.
It is also noteworthy, that AI-support helps the radiologists to rate the radiographs more consistently: while Fleiss' kappa is 0.278 for test A, the agreement between radiologists is higher in test B with a kappa of 0.484, see Table \ref{table:meanRatings}.

\begin{table}[h!]
	\centering
	\captionof{table}{\textbf{Average ratings of radiologists in detecting OA progressive knees}
	}
	\begin{tabular}{cccccc}
		\hline
		\multicolumn{1}{l}{\begin{tabular}[c]{@{}l@{}}Model support\end{tabular}} & \begin{tabular}[c]{@{}c@{}}Sensitivity\\ (95\% CI)\end{tabular} & \begin{tabular}[c]{@{}c@{}}Specificity\\ (95\% CI)\end{tabular} & \begin{tabular}[c]{@{}c@{}}PPV\\ (95\% CI)\end{tabular}       & \begin{tabular}[c]{@{}c@{}}NPV\\ (95\% CI)\end{tabular}       & Fleiss' kappa \\ \hline
		no (test A)                                                                                         & \begin{tabular}[c]{@{}c@{}}0.421\\ (0.275, 0.567)\end{tabular}   & \begin{tabular}[c]{@{}c@{}}0.723\\ (0.592, 0.853)\end{tabular}   & \begin{tabular}[c]{@{}c@{}}0.284\\ (0.244, 0.324)\end{tabular} & \begin{tabular}[c]{@{}c@{}}0.839\\ (0.822, 0.856)\end{tabular} & 0.278         \\ \hline
		yes (test B)                                                                                       & \begin{tabular}[c]{@{}c@{}}0.516\\ (0.413, 0.618)\end{tabular}   & \begin{tabular}[c]{@{}c@{}}0.886\\ (0.791, 0.981)\end{tabular}   & \begin{tabular}[c]{@{}c@{}}0.576\\ (0.460, 0.692)\end{tabular} & \begin{tabular}[c]{@{}c@{}}0.884\\ (0.870, 0.898)\end{tabular} & 0.484         \\ \hline
	\end{tabular}
	\caption*{\small
		In total, 486 knee radiographs were rated by seven radiologists for detecting rapid progressive knees (95 out of 486).
		This reader test consists of two phases: identification of progressors based on baseline radiographs or combining baseline radiograph information together with our model assisted prediction.   
		Fleiss' kappa was computed to represent the inter-reader agreement.
		Abbreviations: PPV: positive predictive value, NPV: negative predictive value.
	}
	\label{table:meanRatings}
\end{table}

\subsection*{Robustness evaluation of our system}
\begin{figure}[h!]
	\centering
	\scalebox{1.0}{
		\includegraphics[trim=40 280 400 10, clip, width=\textwidth]
		{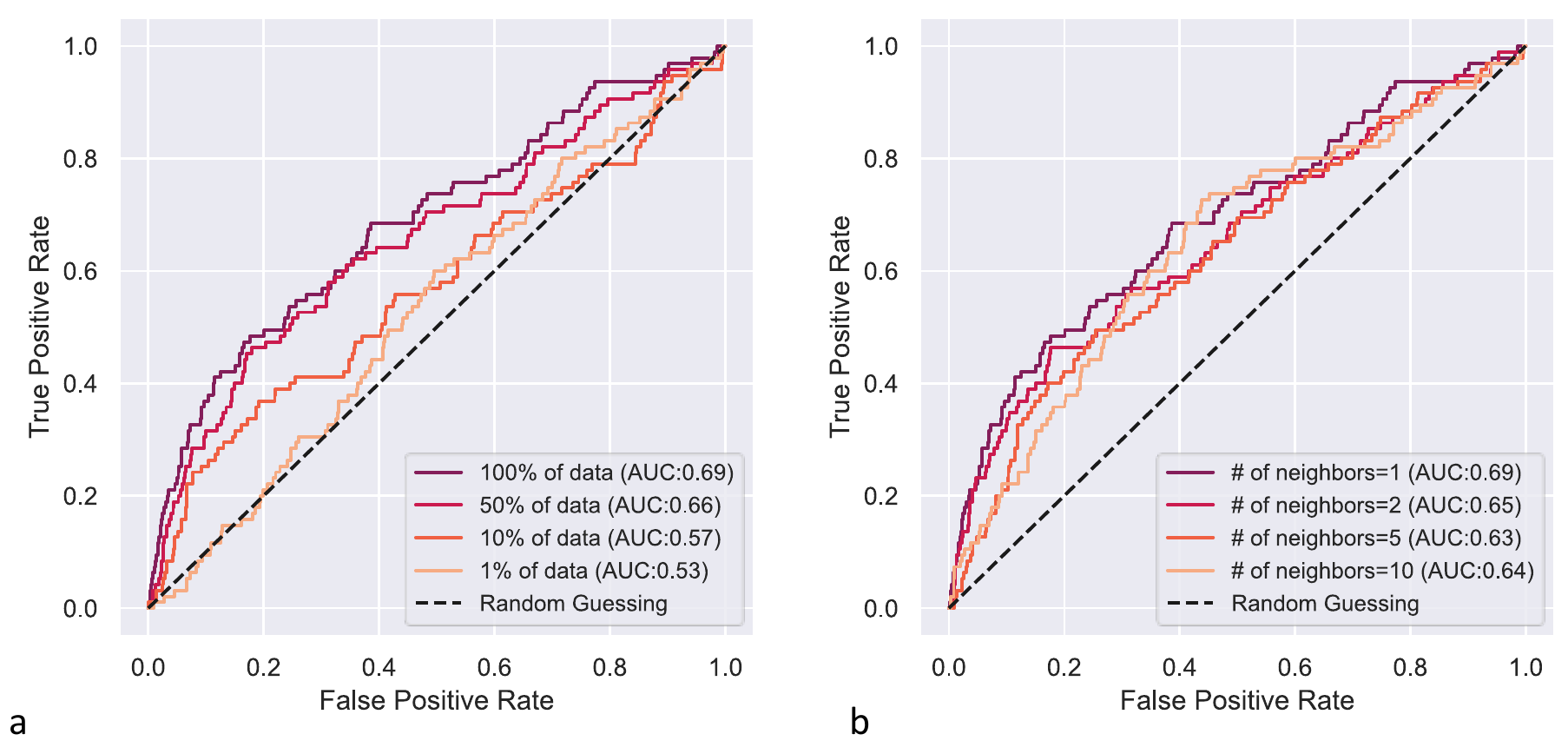}}
	\caption{\label{fig:rob}\textbf{System robustness as a function of training set size and the number of nearest neighbors used.}
		(a) ROC curves with respect to different amounts of training data from the OAI. 
		ROC-AUC was improved by including more data points into the image manifold. 
		(b) Classification performance with respect to variable numbers of nearest neighbors that are utilized to determine the onset or progression of OA. 
		Classification AUC when utilizing 1, 2, 5, or 10 nearest neighbors was comparable (see p-value in Table \ref{Table: ournn}), indicating that adjacent samples in the latent space encode similar information.
	}
\end{figure}

Predicting the future state of a progressive disease via the latent nearest neighbor algorithm relies on two hyperparameters: the size of data that can be embedded into latent space, and the number of nearest neighbors considered (\ref{equ:vectorfield}).
To verify that larger datasets are necessary for accurate predictions, we randomly subsampled the OAI training set to 1\% (443 radiographs), 10\% (4,433 radiographs), and 50\% (22,163 radiographs).
Fig. \ref{fig:rob}a demonstrates that incorporating more data points into the latent space improves the performance (in terms of ROC-AUC) of the proposed method.
Related metrics and confidence intervals (CI) can be found in Supplementary Table \ref{Table: ourd}.
We evaluated the stability of our method with respect to the hyper-parameter $m$ that denotes the number of nearest neighbors taken into account in the generation of future synthetic radiographs.
As demonstrated in Fig. \ref{fig:rob}b and Table \ref{Table: ournn}, only a minor performance change is observed when averaging multiple vectors of the nearest neighbors in the latent space.

\subsection*{Diagnostic information is preserved in the latent space}
\begin{figure}[h!]
	\centering
	\scalebox{1.0}{
		\includegraphics[trim=0 400 170 0, clip, width=\textwidth]
		{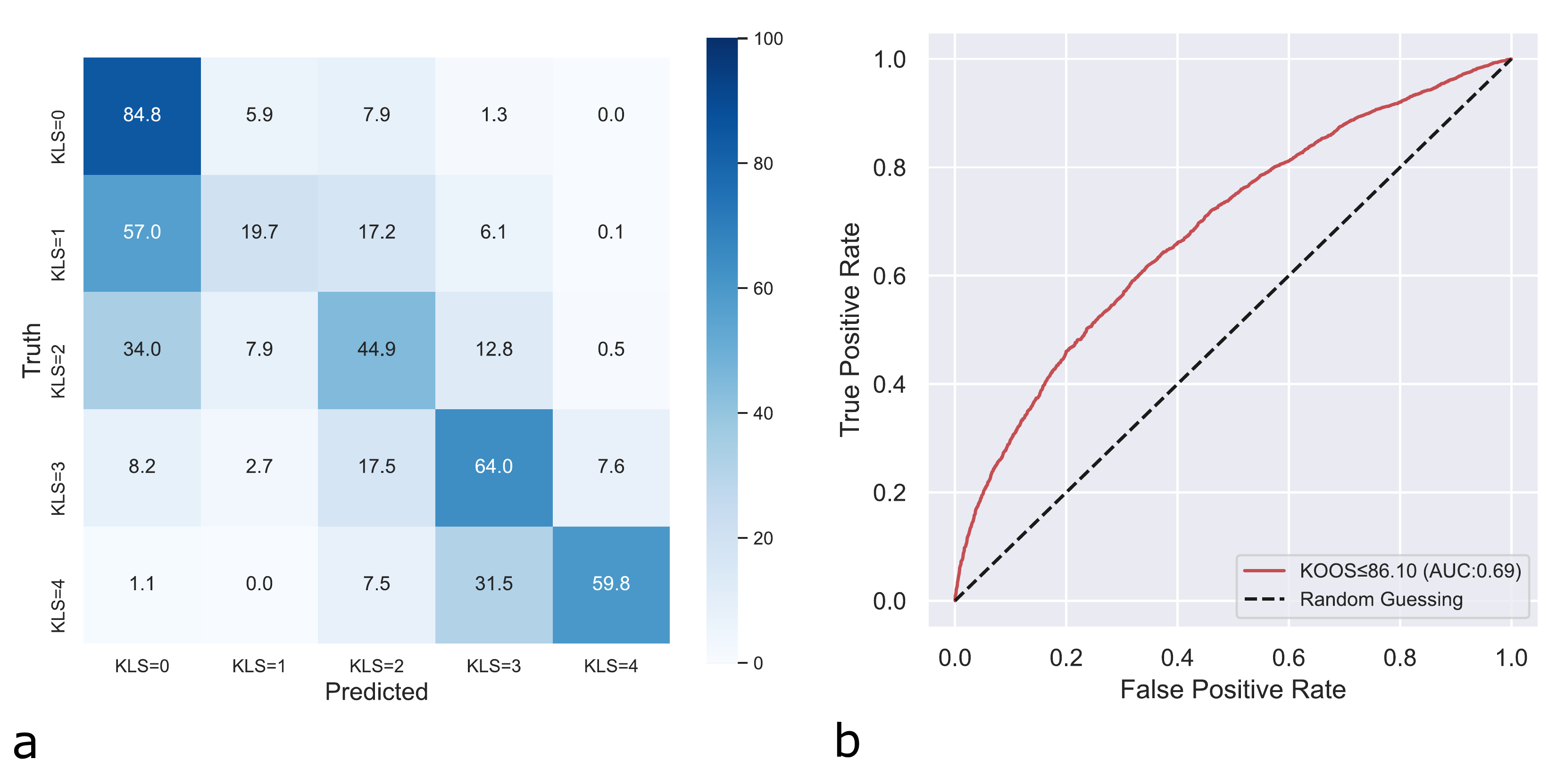}}
	\caption{\label{fig:koos}\textbf{Latent features contain diagnostic information.}
		Exploring the information content of features learned by the predictive model (self-supervised StyleGAN), two classification tasks were performed solely on the low-dimensional latent code: KLS grading classification and severe knee pain classification.
		KLS was correctly predicted as demonstrated by the confusion matrix in (a).	
		In addition, as shown in (b), we trained the shallow classifier on the task of heavy knee pain classification with a threshold of KOOS$\leq$86.1 \cite{englund2003impact}) and reached a state of the art ROC-AUC of 0.69 (95\% CI: 0.651, 0.719) \cite{pierson2021algorithmic}.
	}
\end{figure}

When representing a radiograph solely by a point in the latent space ,i.e., 512-dimensional vector, diagnostic information might get lost.
To test whether this is the case, we conducted additional experiments.
Important diagnostic information contained in the images is the grading according to Kellgren-Lawrence-Score (KLS) and the knee injury and osteoarthritis outcome score (KOOS). 
We followed the approach given in \cite{zhang2016colorful, donahue2019large} and performed the classification task solely on latent vectors - i.e., not the images themselves were used for training, but their low-dimensional condensed representation in latent space.
We trained the algorithm to extract the KLS grade and distinguish severe pain indicated by KOOS $\leq$ 86.1.
Data was partitioned as before (see Fig \ref{fig:datapartition}) and the latent vector for a specific knee was calculated following Algorithm \ref{algo:invert}.
For the classification tasks, we trained a shallow neural network with only one hidden layer on the latent vectors of the OAI training and validation sets.
Results are given in Fig. \ref{fig:koos}a.
When evaluated on the test set, the severe pain classifier reached a state-of-the-art AUC of 0.69 (95\% CI: 0.651, 0.719), a sensitivity of 0.61 (95\% CI: 0.56, 0.66), and a specificity of 0.66 (95\% CI: 0.623, 0.696), on par with algorithms trained on the full images \cite{pierson2021algorithmic}.

\section*{Discussion}
\textcolor{blue}{
In this work, we demonstrated the diagnostic and the prognostic potential of using data without the need for human generated labels to train a machine learning algorithm.}
We focused on knee radiographs, in particular on OA onset and progression as an example.
\textcolor{blue}{
The task of inferring the disease progression from a presymptomatic baseline itself is challenging for machine learning algorithms, and for radiologists (Fig. \ref{fig:roc}d). 
Only less than 4,800 participants were enrolled in the OAI study which leading to a limited amount of baseline radiographs to train an end-to-end model. 
We tried to compensate for this issue using ImageNet weights to initialize the end-to-end learning classifier, however, the benefit is minor. 
Our proposed method generates a synthetic prediction based on both baseline and follow-up scans in the training dataset, e.g., more than 44,000 X-rays are used in self-supervised GAN training.
To better understand the proposed OA progression risk estimation, we plotted the corresponding working pipeline and module structure in Fig \ref{fig:roc}a and b.  
Instead of directly estimating the probability from a single image, we computed the progression risk/probability using a sum of joint probabilities.
When assessing the future progression risk, radiologists focused their attention to joint space narrowing, subchondral sclerosis, the presence of osteophytes and general appearance such as soft tissue around the knee joint. 
All of these factors were clearly visible in the 256$\times$256 radiographs (see Fig. \ref{fig:256vs512}) and the radiologists stated that their rating was not affected by the limited resolution.
}

As OA prognostication remains challenging despite enormous clinical and scientific research efforts, the identification of patients who are susceptible to develop OA or to undergo OA progression is of paramount importance to guide treatment and to potentially facilitate new preventive or curative treatment strategies.
Even though our predictive model has been validated on high-quality study collectives of the OAI and MOST trials and remains to be validated in less homogeneous clinical collectives, the systematic use of a self-supervised generative model without the need for human-generated labels suggests a promising alternative path forward to improve OA prognostication based on predictive models.
Importantly, the presented method can be applied to a plethora of general medical conditions in which the future progression of a disease is important for medical decision making.
No dedicated labeling of data is needed, as the imaging devices produce a timestamp on each acquired radiological image by default.
Synthetic radiological images have been used before to alleviate the problems associated with data sharing \cite{han2020breaking} or to generate synthetic CT datasets from MRI datasets \cite{lei2019mri}.
However, no study so far has involved synthetic radiological images and latent space trajectories into disease prognostication and, secondarily, clinical decision-making.
Possible application scenarios include the use of brain MRI to monitor and predict the onset or progression of Alzheimer's disease or other neuro-degenerative disorders to act early and to prevent further progression \cite{hsu2017primary}.
Indeed, it has been shown that prediction of the progression in patients with Alzheimer's disease based on variables extracted through deep neural networks might be a promising approach \cite{raket2020statistical, louis2019riemannian, lee2019predicting}.
Similar needs arise in the progression of patients with liver fibrosis towards liver cirrhosis \cite{pimpin2018burden}, the progression of osteoporosis \cite{compston2017uk}, the course of cardiovascular diseases \cite{gorenek2017european}, or diabetes and its complications \cite{alkhatib2017functional}.
These application scenarios require big datasets and potential sharing of data across institutions and borders, but this is in reach given the national efforts to implement screening programs \cite{mckinney2020international}.

This study and the predictive model have limitations:
First, to train the generative model a large amount of data is needed - for plain radiographs as used in this work, we utilized approximately 50,000 images.
It can be expected that a higher number of samples will be needed for image data that contain higher dimensional information, e.g., three- or even four-dimensional time-resolved MRI or CT images.
Second, in order for the model to make accurate predictions, a clear progression with corresponding image changes needs to be present.
This is clearly the case for OA and probably for several other diseases that were mentioned above as potential application scenarios.
Previous work has for example focused on predicting brain tumor size \cite{petersen2019deep, petersen2021continuous}.
While these models assumed a Gaussian distribution in latent space, allowing for convenient sampling, the unbounded latent in our work focusses on sample quality to facilitate a hybrid approach for clinical practice.
Note, that our approach can be expanded to quantify uncertainty: we can progressively add more observations to the latent space and measure the information gain.
\textcolor{blue}{
Third, we found in Fig \ref{fig:failure} that sometimes the system correctly predicts the progression of OA, yet allocates progression to the wrong side of the knee.}
Here, the consistent implementation of the model's output, e.g., by saliency maps, may improve explainability and comprehensibility on the part of the user.
Fourth, the OAI and MOST datasets are well curated and provide standardized image quality due to stringently implemented acquisition and post-processing techniques.
The predictive model’s successful application on these datasets does not necessarily translate to clinical applicability and benefit as radiographs acquired in clinical contexts may display larger variability in image quality. 
In conclusion, we introduce and validate a novel predictive model that uses self-supervised learning to provide information on OA onset and progression.
Beyond OA, this proof-of-concept study highlights the potential of self-supervised learning in combination with latent searching schemes, generative models, and latent space trajectories to make substantiated predictions on future disease states in nearly all fields of medicine without necessitating any additional labels.


\clearpage
\section*{Materials and methods}
\subsection*{Development and testing datasets}
The system was developed on data from the Osteoarthritis Initiative (OAI), consisting of a total number of 26,525 bilateral fixed flexion plain film radiographs from 4,796 participants with OA of the knee or at risk of developing knee OA \cite{eckstein2012recent}. 
In this longitudinal cohort study, the imaging, biochemical, and constitutional findings associated with knee OA have been investigated and documented from February 2004 to October 2015. 
Briefly, the right and the left knees of participants were imaged together \textcolor{blue}{and later stitched to digital radiographs}.
In total, we utilized a pre-trained Hourglass network and extracted 52,981 (separated left and right) knee radiographs with a region of 140mm $\times$ 140mm from all participants who were regularly imaged from baseline to the 96-month follow-up \cite{tiulpin2019kneel}. 
During the development of this study, the input knee images were flipped to the right knee configuration and resized to 256 $\times$ 256 by bilinear \textcolor{blue}{decimation}.
Both quantitative and semi-quantitative assessments, such as the joint space width and the KLS, were available for most radiographs (43,132 knee radiographs from 3,944 patients) from the OAI. 
In this study, we used the KLS as a quantitative measure image's morphological changes reflective of OA onset or progression.
As shown in Fig \ref{fig:datapartition}, participants were randomly allocated to training, validation, and hold-out test sets.
The test set was kept separate during the development phase of the algorithm and was only used to test the algorithm's performance after parameter tuning had been finished.
Since the KLS is not needed in the training of our GAN, the unlabeled radiographs could be used as well and were included in the model development loop (9,849 radiographs from 852 patients).
Supplementary Table \ref{Table: xrays_meta}, describes demographic characteristics and the KLS for relevant subsets of the OAI dataset.

An independent dataset from the Multi-center Osteoarthritis Study (MOST) \cite{segal2013multicenter} was also used to validate the system's performance and generalization in another dataset from another trial.
Similar to the OAI dataset, the MOST dataset is a longitudinal, prospective study that recruited more than 3,026 participants for five follow-up visits at 15, 30, 60, 72 and 84 months. 
In the MOST dataset, all 19,340 knee radiographs labeled with KLS (bilateral, standing fixed-flexed posteroanterior views) were used solely for testing our proposed predictive model. 
Detailed label statistics for the MOST datasets can be found in table \ref{Table: xrays_meta}.
Ethical approval for collecting all subject information was provided by the OAI and MOST.

\subsection*{Details of system modules}
In general, our system is trained to take the baseline knee radiograph and automatically produce the future knee morphology as well as the risk score predicting whether the patient will suffer from OA onset or progression in the next years.
The proposed modular approach consists of a generative StyleGAN, a latent back-mapping module, a synthetic future radiograph module, and a progression risk estimation module. 

\subsubsection*{StyleGAN and training details}
Our approach relies on training a generative model that learns high-level representations of input knee radiographs.
A GAN consists of a generator network and a discriminator network.
The generator network G projects a simple latent distribution $P_z$ to a more complex data distribution $P_{g}$. 
Next, a feed forward neural network, i.e., a discriminator D, is used to directly measure the distributional distance between $P_{g}$ and a real data distribution $P_{r}$.
The training process of our GAN  follows the original approach by solving the min-max problem $\underset{G}{\min} \, \underset{D}{\max}\;L(G, \, D)$ via training D and G alternately.
The objective function of the GAN can be expressed as:
\begin{equation}
	\begin{split}
	\displaystyle L_D &= \min_{D}\underset{x \sim P_r(x)}{\mathbb{E}}[-\text{log}\,D(x)]+\underset{x\sim P_g(x)}{\mathbb{E}}[-\text{log}(1 - D(x))] \\
	\displaystyle L_G &= \min_{G}\underset{z \sim P_z(z)}{\mathbb{E}}[-\text{log}\,D(G(z))].
	\end{split}
	\label{equ:ganloss}
\end{equation}
Goodfellow et al. suggest this version of the G loss $L_G$ in equation \ref{equ:ganloss} to circumvent the saturation of G during the early phase of training.
However, plenty of observations have been made on problems such as the instability of training a GAN to truly converge and the mode collapse issue in GAN-generated samples. 
It has been widely shown that optimizing the objective in equation \ref{equ:ganloss} via simultaneous gradient descent does not guarantee the convergence towards the Nash equilibrium point \cite{goodfellow2014generative}. 
Therefore, it's necessary to add additional regularization strategies to stabilize the training. 
In this work, we regularize the training process of D $\theta_d$ via adding a gradient penalty term $R_1$ to $L_D$:  
\begin{equation}
	\displaystyle R_1(\theta_d) = \frac{\gamma}{2}\;\underset{x \sim p_r(x)}{\mathbb{E}}\left \| \nabla_x\,D_{\theta_d} (x) \right \|^2.
	\label{equ:l1_regularization}
\end{equation}
We set the $R_1$ regularization weighting factor to $\gamma=10$ throughout the training process.
Similarly, we enforce the smoothness of the G function $\theta_g$ with respect to the intermediate latent $w$ by using the path length regularization \cite{karras2020analyzing}: 
\begin{equation}
	R_{pl}(\theta_g) = \underset{w,\, y \sim \mathcal{N}(0, I)}{\mathbb{E}}(\left \| \nabla_{w}\,G(w) y \right \|_2 - a)^2,
	\label{equ:lpl_regularization}
\end{equation}
where $y$ is randomly distributed Gaussian noise with the same shape as the generated image $G(w)$ and the constant $a$ is the running average of $\underset{w,\, y}{\mathbb{E}}\left \| \nabla_{w}\,G(w) y \right \|_2$. 
This G regularization is of particular importance to our method as it guarantees meaningful interpolations in the intermediate latent space and makes the G inversion easier \cite{karras2020analyzing}. 
Figure \ref{fig:GAN_train} a plots the output logits for both real and generated samples from D during training.
Regularization on both D and G are also plotted in Figure \ref{fig:GAN_train} b.

To measure whether the generated images converged to realistic radiographs, we continuously measured the Fréchet Inception Distance (FID) between a set of 50,000 real and synthesized pairs during training \cite{heusel2017gans}.
We found that the FID decreased nearly monotonically, indicating that the general appearance of the generated images approaches that of real radiographs.
The corresponding figures depicting the evolution of the FID are given in Fig. \ref{fig:GAN_train} c.
However, a single FID score cannot make conclusions about the generation ability, i.e., precision and recall, of the trained GAN \cite{sajjadi2018assessing}.  
We additionally computed the distribution-wise precision and recall parameters to characterize our GAN and to pick the best performing model. 
As shown in Figure \ref{fig:GAN_train} d, we selected the best performing GAN model with the highest precision and recall (indicated by the red arrow).
The description of the selected model characteristics to the target distribution of knee radiographs is depicted in a precision-recall curve \cite{sajjadi2018assessing} in Figure \ref{fig:GAN_train} e.

We employ the improved G architecture of StyleGAN \cite{karras2020analyzing}.
Unlike conventional generators, Karras et al. design additional mapping layers $f$ (two-layer perceptron) that non-linearly project the simple Gaussian latent $z$ into an intermediate latent space $W$ with the same 512 dimensions. 
It has been shown that feature disentanglement becomes achievable in the intermediate latent space $W$ as it does not have to be bounded by any fixed, e.g., Gaussian or uniform, distribution for convenient sampling. 
In this study, we therefore refer to the latent vector as vectors $w$ in $W$.
To control feature appearance, as shown in Fig. \ref{fig:GAN_nn} c-e, we pass the affine transformed latent vector w to the convolution layers of the StyleGAN G \cite{karras2020analyzing}.
The G architecture resulted in a final spatial resolution of 256$\times$256 is shown in Fig. \ref{fig:GAN_nn}.
We picked the leaky rectified linear unit (LReLU) with a negative slope of 0.2 and weight demodulation \cite{karras2020analyzing} as the activation function and normalization in G layers. 
Similarly, in D, we chose LReLU and Sigmoid as activation functions for convolution layers and the final classifier layer separately. 
During training, we used two Adam optimizers \cite{kingma2014adam} with the same learning rate of 0.0025 for both G and D, and the mini-batch size was set to 32 across two NVIDIA Titan RTX GPUs.
Throughout the study, Pytorch v.1.8.1 was used for implementation \cite{paszke2019pytorch}.

\subsubsection*{Latent back mapping module}
Once the generative model is trained, a method is needed to map an existing knee radiograph $x$ to a point $w$ in latent space $W$ such that the radiograph $\hat x$ synthesized from $w$ is a close representation of $x$.
Technically, under the framework of StyleGAN, we need to optimize the intermediate latent vector $w$ and noise maps $n_i$ corresponding to different resolutions for reconstructing the counterpart of a target image $x$.
To approximate the mapping function $\text{G}^{-1}(x)$, we compute the learned perceptual image patch similarity (LPIPS) \cite{zhang2018translating}, i.e., the Euclidean distance between embeddings of $x$ and $\hat x$ extracted by a pretrained VGG-16 network, as the cost function (Fig. \ref{fig:inverseG} a).  
While optimizing $n_i$, we make sure no meaningful gradient from $x$ is leaked to $n_i$ during gradient descent by adding an auto-correlation regularization \cite{karras2020analyzing}. 
We additionally impose randomization to the latent $w$ when evaluating the cost function in order to avoid trapping in a local minimum \cite{lipton2017precise}.  
The cost was optimized by an Adam optimizer with an initial learning rate of 0.1.
More details of the inverting a StyleGAN G are depicted in Algorithm \ref{algo:invert}.
To test, whether the computed $w$ is a good approximation of $\text{G}^{-1}(x)$, we computed the structural similarity index (SSIM) between 5,000 pairs of original and reconstructed radiograph  of both the OAI and the MOST dataset in Fig. \ref{fig:inverseG} b. 
As shown in Fig. \ref{fig:inverseG} b, we observed that OAI reconstructed radiographs achieved SSIMs of around 0.8, while MOST reconstructed radiographs displayed slightly larger variability, most likely due to the fact that the GAN had been trained on OAI radiographs only.
\textcolor{blue}{
In addition, 500 pairs of $x$ and their corresponding $G(w)$ were also compared by one radiologist with 8 years of experience and checked visually for inconsistencies on a scale of 0 (the synthesized image can clearly be identified as artificial due to artefacts) to 5 (the synthesized image is not discernable from the real image).
The mean score in this evaluation was $4.2\pm 0.4$ underlining the good agreement between real and synthesized images.
}

\begin{algorithm}[h]
	\caption{\textbf{Inverting generator} $\text{G}^{-1}(x)$. We use default values of $\eta$ = 0.1, $\alpha$ = $10^5$, and $n_z$ = 10,000.}\label{algo:invert}
	\begin{algorithmic}[1]
		\State \textbf{Require:} A candidate baseline knee radiograph $I$; a pretrained StyleGAN $G(w, n_i)$ with a mapping network $f(z)$; LPIPS loss function $L_{\text{LPIPS}}$; noise regularization $R$; learning rate schedule $T$.
		\State \textbf{Require:} a maximum learning rate $\eta$; the number of latent code $z$ sampled $n_z$; the weight of noise regularization $\alpha$; a constant $t$ scales the noise amplitude of $w$.
		\State \textbf{Output:} the optimized latent vector $w$ that can reconstruct $I$. 
		\State \For {$n = 1, ..., n_z$}
		\State Sample normally distributed latent code $z_n$
		\EndFor
		\State Compute  $ \mu_{w} = \mathbb{E}_n\; f(z_n)$ and $ \sigma_{w}^2 = \mathbb{E}_n\; \left \| f(z_n)-\mu_{w} \right \|_2^2$
		\State Initialize latent $w=\mu_{w}$ and noise map $n_i = \mathcal{N}(0, \textbf{I})$
		\State \While {not converged}
		\State $w \gets w + \mathcal{N}(0,\, 0.05\sigma_{w}\,t)$
		\State $L \gets L_{\text{LPIPS}}(I, G(w, n_i)) + \alpha R(n_i)$
		\State $w \gets w - \eta\nabla_wL$
		\State $n_i \gets n_i - \eta\nabla_{n_i}L$
		\State $\eta \gets T(\eta)$
		\EndWhile
		\State \Return $w$
	\end{algorithmic}
\end{algorithm}

\subsubsection*{Synthetic future radiograph generation} \label{char: futuregen}
In the first step, we construct a dictionary $D=\{w_i\}_{i=1}^N$ that contains mapped latent code $w$ of OAI radiographs based on Algorithm \ref{algo:invert}.
We conceptually construct a vector field $\bf{V}$ that maps the points in the latent space $W$ that belong to the baseline radiographs $x^{\text{\,baseline}}$ to points in the latent space of the follow-up radiograph $x^{\text{\,follow-up}}$: $\textbf{V}: \text{G}^{-1}(x^{\text{\,baseline}}) \rightarrow \text{G}^{-1}(x^{\text{\,follow-up}})$.
Predicting the future state of a progressive knee $x$ during inference is a combination of two steps: identification of the latent code $w_{x}$ of the candidate's knee (Algorithm \ref{algo:invert}), and modeling the future appearance of the knee radiograph via interpolating the latent vector field $V$.  
To interpolate the value of the vector field at new positions $x$, we employ the nearest neighbor interpolation with \textit{nearest} defined via the normalized cosine distance
\begin{equation}
	\displaystyle \text{d}(x, y) = \left \| \frac{w_x}{\left \| w_x \right \|_2} - \frac{w_y}{\left \| w_y \right \|_2} \right \|_2,
	\label{equ:distance}
\end{equation}

and the actual interpolation defined as:
\begin{equation}
	\displaystyle \Delta w^{nn}_x = \frac{1}{m} \sum_{i=1}^{m}\frac{\Delta \, t^i}{\Delta \, T}\left[ \text{G}^{-1}(x_{\,i}^{\text{\,follow-up}})-\text{G}^{-1}(x_{\,i}^{\text{\,baseline}}) \right],
	\label{equ:vectorfield}
\end{equation}
where $m$ is the number of nearest neighbors, $\Delta T$ is the time interval denoting how far in the future the synthesized radiograph should be generated, and $\text{G}^{-1}(x_{\,i})$ is the corresponding latent vector of the i-th nearest neighbor of $x$.
The prediction of the future knee state can be obtained via feeding the mapped latent $w_{\hat x}$ and the computed vector difference $\Delta w^{nn}$ to the generator, i.e., $G(w_{x}+\Delta w^{nn}_x)$.
This procedure was used to generate synthetic future radiographs for the human reader test.
\begin{algorithm}[h]
	\caption{\textbf{Radiograph prediction as generated from the latent space}}\label{algo:future}
	\begin{algorithmic}[1]
		\State \textbf{Require:} A candidate baseline knee radiograph $I$; a pretrained StyleGAN $G$; the calculated mapping dictionary $D$ of knees.
		\State \textbf{Require:} the number of nearest neighbors considered $m$.
		\State \textbf{Output:} Synthetic future radiograph of the candidate's knee.
		\State get $w_I \gets \text{G}^{-1}(I)$ from Algorithm \ref{algo:invert} 
		\State \For {$w_j$ in $D.keys()$} 
		\State $d(w_I, \, w_j) \gets \left \| \cfrac{w_I}{\left \| w_I \right \|_2} - \cfrac{w_j}{\left \| w_j \right \|_2} \right \|_2$ 
		\EndFor
		\State Sort $w_j$ in $D.keys()$ according to distances $d(w_I, \, w_j)$
		\State Initialize $\Delta w^{nn}_I=0$
		\State \For {$w_i$ in set $M$ containing latents for the $m$ smallest distances $d(w_I, \, w_j)$}
		\State $\Delta w^{nn}_I \gets \frac{\Delta \, t_i}{\Delta \, T}(w_i^{follow-up} - w_i^{baseline}) + \Delta w^{nn}_I$
		\EndFor
		\State \Return $I_{\text{future}} \gets G(w_I + \frac{1}{m}\Delta w^{nn}_I,\; n_i=0)$
	\end{algorithmic}
\end{algorithm}

\subsubsection*{Progression risk estimation module}
The risk prediction module takes in a baseline scan and its (synthesized) follow-up scan in the form of gray-scale knee radiographs to predict the OA progression risk. 
Previous studies solely utilized the baseline patient state, e.g., the radiological image or study, and tried to predict the future progression probability in an end-to-end fashion \cite{tiulpin2019multimodal, yim2020predicting, googlenaturelungcancer}. 
However, in this study, we can predict the morphological appearance of the radiograph at a future time point (Algorithm \ref{algo:future}) and compute the risk based on the appearance of the synthetic radiograph as detailed above. 
Inside the module, we placed an adversarially trained ResNet-50 \cite{han2021advancing} that can correctly classify the KLS of the input knee radiograph. 
As indicated in Fig. \ref{fig:datapartition}, the classifier was developed and tested on 34,477 KLS-labeled knee radiographs from 2,654 participants and another 8,655 knee radiographs from 1,290 participants that were initially held out as the radiologists’ test set but were then included.
The ResNet classifier utilized the softmax loss function as its criterion.
Additionally, to improve the interpretability of the classifier, we applied adversarial augmentation with an attack strength of $\epsilon = 0.005$ during the training process \cite{han2021advancing}. 
The classification performance of the ResNet, as well as its interpretable gradient saliency maps are shown in Fig. \ref{fig:roc} a and b.

We denote $x_i$, $x_j$ as baseline and follow-up knee radiographs respectively (Fig. \ref{fig:overview}c), and  $c_i$, $c_j$ as KLS classes for $x_i$ and $x_j$ ranging from 0 to 4.
We defined patients with imminent OA onset and/or progression towards osteoarthritis changes as those, who showed progress of more than one KLS ($c_j - c_i > 1$) over scans.
We then computed the probability of OA progression ($y=1$) as a sum of joint probabilities $\{ (c_i,\, c_j) \}$ which fulfill the condition that the KLS change is larger than 1, i.e., $c_j - c_i > 1$, between its prior and follow-up visit radiographs:
\begin{equation}
	\begin{aligned}
		p(y=1|x_i, x_j) &= \sum_{\{ (c_j-c_i > 1) \}}p(\text{KLS}=c_i|x_i)\times p(\text{KLS}=c_j|x_j), \\
	 p(y=0|x_i, x_j) &=  \sum_{\{ (c_j-c_i \leq 1) \}}p(\text{KLS}=c_i|x_i)\times p(\text{KLS}=c_j|x_j).
	\end{aligned}
	\label{equ:joint}
\end{equation}

\subsection*{Reader experiments}
To identify patients who are in need of preventive care, it is important to differentiate between patient with imminent OA onset and/or progression towards osteoarthritic changes and patients with stable disease.
We defined participants belonging to the group of  progressors as those, who showed progress of more than one KLS over 8 years.
We designed two different experiments:
In experiment A, seven experienced radiologists were given only the baseline radiograph and were asked to predict whether a patient will experience the onset or progression of OA, or whether a patient exhibits a stable disease.
In experiment B, the proposed generative model was used to synthesize the most probable imaging appearance (see section \nameref{char: futuregen}) of the radiograph 8 years into the future based on the vector field in the latent space.
The radiologists were then given the predicted synthesized radiograph next to the baseline radiograph as additional guiding help and were again asked to perform the same allocation into two groups.
For both experiments, we randomly selected 468 participants and 486 knee radiographs (95 progressive as defined above) with a resolution of $256\times256$ out of the test set. 
All reading experiments were done on diagnostic computer monitors. 

\subsection*{Statistical analysis}
For each of the experiments, we calculated the following parameters on the test set: ROC-AUC, sensitivity, and specificity.
The cutoff value for deciding between the presence or non-presence of pathology was determined by minimizing  $(1- \text{sensitivity})^2 + (1- \text{specificity})^2$ \cite{kniep2019radiomics}.
If not otherwise stated, standard deviations (SD) and 95\% CI were extracted using bootstrapping with 10,000 redraws.
The difference in metrics, such as ROC-AUC, sensitivity, and specificity, was defined as a $\Delta$-metric. 
For the total number of N = 1,000 bootstrapping, models were built after randomly permuting predictions of two classifiers, and metric differences $\Delta - \text{metric}_i$ were computed from their respective scores. 
We obtained the two-tailed p-value of individual metrics by counting all $\Delta - \text{metric}_i$ above the threshold $\Delta$metric.
Statistical significance was defined as P $<$ 0.001.
For the clinical reader experiments, Cohen's kappa was used to calculate intra-reader agreement between experiments A and B for the same reader, while Fleiss' kappa was used to calculate inter-reader agreement between the seven radiologists.
\\ \\
\textbf{Author Contributions:}
TH and DT devised the concept of the study, DT, SN, MS, MZ, FP, SK, and MH performed the reader tests. TH wrote the code and performed the performance studies. TH and DT did the statistical analysis. TH and DT wrote the first draft of the manuscript. All authors contributed to correcting the manuscript.
\textbf{Disclosures of Conflicts of Interest:}
JNK declares consulting services for Owkin, France and Panakeia, UK.
\textbf{Funding Sources:} 
JNK is supported by the German Federal Ministry of Health 
(DEEP LIVER, ZMVI1-2520DAT111) and the Max-Eder-Programme of the German Cancer Aid (grant 70113864).
\textbf{Data and Materials Availability:} 
All datasets used in this study are publicly available:
the dataset of knee radiographs of the OAI and MOST datasets can be requested from \url{https://nda.nih.gov/oai/} and \url{https://agingresearchbiobank.nia.nih.gov/studies/most/}.
The code used in this study is made fully publicly available under \url{https://github.com/peterhan91/disease_progression}.

\bibliography{sources.bib}

\begin{thebibliography}{10}

\bibitem{grigorescu2020survey}
S.~Grigorescu, B.~Trasnea, T.~Cocias, and G.~Macesanu, ``A survey of deep
  learning techniques for autonomous driving,'' {\em Journal of Field
  Robotics}, vol.~37, no.~3, pp.~362--386, 2020.

\bibitem{krizhevsky2012imagenet}
A.~Krizhevsky, I.~Sutskever, and G.~E. Hinton, ``Imagenet classification with
  deep convolutional neural networks,'' {\em Advances in neural information
  processing systems}, vol.~25, pp.~1097--1105, 2012.

\bibitem{senior2020improved}
A.~W. Senior, R.~Evans, J.~Jumper, J.~Kirkpatrick, L.~Sifre, T.~Green, C.~Qin,
  A.~{\v{Z}}{\'\i}dek, A.~W. Nelson, A.~Bridgland, {\em et~al.}, ``Improved
  protein structure prediction using potentials from deep learning,'' {\em
  Nature}, vol.~577, no.~7792, pp.~706--710, 2020.

\bibitem{jumper2021highly}
J.~Jumper, R.~Evans, A.~Pritzel, T.~Green, M.~Figurnov, O.~Ronneberger,
  K.~Tunyasuvunakool, R.~Bates, A.~{\v{Z}}{\'\i}dek, A.~Potapenko, {\em
  et~al.}, ``Highly accurate protein structure prediction with alphafold,''
  {\em Nature}, vol.~596, no.~7873, pp.~583--589, 2021.

\bibitem{lundberg2018explainable}
S.~M. Lundberg, B.~Nair, M.~S. Vavilala, M.~Horibe, M.~J. Eisses, T.~Adams,
  D.~E. Liston, D.~K.-W. Low, S.-F. Newman, J.~Kim, {\em et~al.}, ``Explainable
  machine-learning predictions for the prevention of hypoxaemia during
  surgery,'' {\em Nature biomedical engineering}, vol.~2, no.~10, pp.~749--760,
  2018.

\bibitem{recht2018cifar}
B.~Recht, R.~Roelofs, L.~Schmidt, and V.~Shankar, ``Do cifar-10 classifiers
  generalize to cifar-10?,'' {\em arXiv preprint arXiv:1806.00451}, 2018.

\bibitem{deng2009imagenet}
J.~Deng, W.~Dong, R.~Socher, L.-J. Li, K.~Li, and L.~Fei-Fei, ``Imagenet: A
  large-scale hierarchical image database,'' in {\em 2009 IEEE conference on
  computer vision and pattern recognition}, pp.~248--255, Ieee, 2009.

\bibitem{petersen2010alzheimer}
R.~C. Petersen, P.~Aisen, L.~A. Beckett, M.~Donohue, A.~Gamst, D.~J. Harvey,
  C.~Jack, W.~Jagust, L.~Shaw, A.~Toga, {\em et~al.}, ``Alzheimer's disease
  neuroimaging initiative (adni): clinical characterization,'' {\em Neurology},
  vol.~74, no.~3, pp.~201--209, 2010.

\bibitem{menze2014multimodal}
B.~H. Menze, A.~Jakab, S.~Bauer, J.~Kalpathy-Cramer, K.~Farahani, J.~Kirby,
  Y.~Burren, N.~Porz, J.~Slotboom, R.~Wiest, {\em et~al.}, ``The multimodal
  brain tumor image segmentation benchmark (brats),'' {\em IEEE transactions on
  medical imaging}, vol.~34, no.~10, pp.~1993--2024, 2014.

\bibitem{casey2018adolescent}
B.~Casey, T.~Cannonier, M.~I. Conley, A.~O. Cohen, D.~M. Barch, M.~M. Heitzeg,
  M.~E. Soules, T.~Teslovich, D.~V. Dellarco, H.~Garavan, {\em et~al.}, ``The
  adolescent brain cognitive development (abcd) study: imaging acquisition
  across 21 sites,'' {\em Developmental cognitive neuroscience}, vol.~32,
  pp.~43--54, 2018.

\bibitem{eckstein2012recent}
F.~Eckstein, W.~Wirth, and M.~C. Nevitt, ``Recent advances in osteoarthritis
  imaging—the osteoarthritis initiative,'' {\em Nature Reviews Rheumatology},
  vol.~8, no.~10, pp.~622--630, 2012.

\bibitem{yim2020predicting}
J.~Yim, R.~Chopra, T.~Spitz, J.~Winkens, A.~Obika, C.~Kelly, H.~Askham,
  M.~Lukic, J.~Huemer, K.~Fasler, {\em et~al.}, ``Predicting conversion to wet
  age-related macular degeneration using deep learning,'' {\em Nature
  Medicine}, vol.~26, no.~6, pp.~892--899, 2020.

\bibitem{bijlsma2011osteoarthritis}
J.~W. Bijlsma, F.~Berenbaum, and F.~P. Lafeber, ``Osteoarthritis: an update
  with relevance for clinical practice,'' {\em The Lancet}, vol.~377, no.~9783,
  pp.~2115--2126, 2011.

\bibitem{Martel-Pelletier2016}
J.~Martel-Pelletier, A.~J. Barr, F.~M. Cicuttini, P.~G. Conaghan, C.~Cooper,
  M.~B. Goldring, S.~R. Goldring, G.~Jones, A.~J. Teichtahl, and J.-P.
  Pelletier, ``Osteoarthritis,'' {\em Nature Reviews Disease Primers}, vol.~2,
  p.~16072, Oct 2016.

\bibitem{conaghan2013osteoarthritis}
P.~G. Conaghan, ``Osteoarthritis in 2012: parallel evolution of oa phenotypes
  and therapies,'' {\em Nature Reviews Rheumatology}, vol.~9, no.~2, p.~68,
  2013.

\bibitem{arden2006osteoarthritis}
N.~Arden and M.~C. Nevitt, ``Osteoarthritis: epidemiology,'' {\em Best practice
  \& research Clinical rheumatology}, vol.~20, no.~1, pp.~3--25, 2006.

\bibitem{goodfellow2014generative}
I.~Goodfellow, J.~Pouget-Abadie, M.~Mirza, B.~Xu, D.~Warde-Farley, S.~Ozair,
  A.~Courville, and Y.~Bengio, ``Generative adversarial nets,'' {\em Advances
  in neural information processing systems}, vol.~27, 2014.

\bibitem{karras2019style}
T.~Karras, S.~Laine, and T.~Aila, ``A style-based generator architecture for
  generative adversarial networks,'' in {\em Proceedings of the IEEE/CVF
  Conference on Computer Vision and Pattern Recognition}, pp.~4401--4410, 2019.

\bibitem{karras2020analyzing}
T.~Karras, S.~Laine, M.~Aittala, J.~Hellsten, J.~Lehtinen, and T.~Aila,
  ``Analyzing and improving the image quality of stylegan,'' in {\em
  Proceedings of the IEEE/CVF Conference on Computer Vision and Pattern
  Recognition}, pp.~8110--8119, 2020.

\bibitem{han2021advancing}
T.~Han, S.~Nebelung, F.~Pedersoli, M.~Zimmermann, M.~Schulze-Hagen, M.~Ho,
  C.~Haarburger, F.~Kiessling, C.~Kuhl, V.~Schulz, {\em et~al.}, ``Advancing
  diagnostic performance and clinical usability of neural networks via
  adversarial training and dual batch normalization,'' {\em Nature
  Communications}, vol.~12, no.~1, pp.~1--11, 2021.

\bibitem{tsipras2018robustness}
D.~Tsipras, S.~Santurkar, L.~Engstrom, A.~Turner, and A.~Madry, ``Robustness
  may be at odds with accuracy,'' {\em arXiv preprint arXiv:1805.12152}, 2018.

\bibitem{zhang2016colorful}
R.~Zhang, P.~Isola, and A.~A. Efros, ``Colorful image colorization,'' in {\em
  European conference on computer vision}, pp.~649--666, Springer, 2016.

\bibitem{donahue2019large}
J.~Donahue and K.~Simonyan, ``Large scale adversarial representation
  learning,'' {\em arXiv preprint arXiv:1907.02544}, 2019.

\bibitem{pierson2021algorithmic}
E.~Pierson, D.~M. Cutler, J.~Leskovec, S.~Mullainathan, and Z.~Obermeyer, ``An
  algorithmic approach to reducing unexplained pain disparities in underserved
  populations,'' {\em Nature Medicine}, vol.~27, no.~1, pp.~136--140, 2021.

\bibitem{englund2003impact}
M.~Englund, E.~M. Roos, and L.~Lohmander, ``Impact of type of meniscal tear on
  radiographic and symptomatic knee osteoarthritis: a sixteen-year followup of
  meniscectomy with matched controls,'' {\em Arthritis \& Rheumatism: Official
  Journal of the American College of Rheumatology}, vol.~48, no.~8,
  pp.~2178--2187, 2003.

\bibitem{han2020breaking}
T.~Han, S.~Nebelung, C.~Haarburger, N.~Horst, S.~Reinartz, D.~Merhof,
  F.~Kiessling, V.~Schulz, and D.~Truhn, ``Breaking medical data sharing
  boundaries by using synthesized radiographs,'' {\em Science Advances},
  vol.~6, no.~49, p.~eabb7973, 2020.

\bibitem{lei2019mri}
Y.~Lei, J.~Harms, T.~Wang, Y.~Liu, H.-K. Shu, A.~B. Jani, W.~J. Curran, H.~Mao,
  T.~Liu, and X.~Yang, ``Mri-only based synthetic ct generation using dense
  cycle consistent generative adversarial networks,'' {\em Medical physics},
  vol.~46, no.~8, pp.~3565--3581, 2019.

\bibitem{hsu2017primary}
D.~Hsu and G.~A~Marshall, ``Primary and secondary prevention trials in
  alzheimer disease: looking back, moving forward,'' {\em Current Alzheimer
  Research}, vol.~14, no.~4, pp.~426--440, 2017.

\bibitem{raket2020statistical}
L.~L. Raket, ``Statistical disease progression modeling in alzheimer disease,''
  {\em Frontiers in big Data}, vol.~3, 2020.

\bibitem{louis2019riemannian}
M.~Louis, R.~Couronn{\'e}, I.~Koval, B.~Charlier, and S.~Durrleman,
  ``Riemannian geometry learning for disease progression modelling,'' in {\em
  International Conference on Information Processing in Medical Imaging},
  pp.~542--553, Springer, 2019.

\bibitem{lee2019predicting}
G.~Lee, K.~Nho, B.~Kang, K.-A. Sohn, and D.~Kim, ``Predicting alzheimer’s
  disease progression using multi-modal deep learning approach,'' {\em
  Scientific reports}, vol.~9, no.~1, pp.~1--12, 2019.

\bibitem{pimpin2018burden}
L.~Pimpin, H.~Cortez-Pinto, F.~Negro, E.~Corbould, J.~V. Lazarus, L.~Webber,
  N.~Sheron, E.~H.~S. Committee, {\em et~al.}, ``Burden of liver disease in
  europe: epidemiology and analysis of risk factors to identify prevention
  policies,'' {\em Journal of hepatology}, vol.~69, no.~3, pp.~718--735, 2018.

\bibitem{compston2017uk}
J.~Compston, A.~Cooper, C.~Cooper, N.~Gittoes, C.~Gregson, N.~Harvey, S.~Hope,
  J.~Kanis, E.~McCloskey, K.~E. Poole, {\em et~al.}, ``Uk clinical guideline
  for the prevention and treatment of osteoporosis,'' {\em Archives of
  osteoporosis}, vol.~12, no.~1, p.~43, 2017.

\bibitem{gorenek2017european}
B.~Gorenek, A.~Pelliccia, E.~J. Benjamin, G.~Boriani, H.~J. Crijns, R.~I.
  Fogel, I.~C. Van~Gelder, M.~Halle, G.~Kudaiberdieva, D.~A. Lane, {\em
  et~al.}, ``European heart rhythm association (ehra)/european association of
  cardiovascular prevention and rehabilitation (eacpr) position paper on how to
  prevent atrial fibrillation endorsed by the heart rhythm society (hrs) and
  asia pacific heart rhythm society (aphrs),'' {\em European journal of
  preventive cardiology}, vol.~24, no.~1, pp.~4--40, 2017.

\bibitem{alkhatib2017functional}
A.~Alkhatib, C.~Tsang, A.~Tiss, T.~Bahorun, H.~Arefanian, R.~Barake, A.~Khadir,
  and J.~Tuomilehto, ``Functional foods and lifestyle approaches for diabetes
  prevention and management,'' {\em Nutrients}, vol.~9, no.~12, p.~1310, 2017.

\bibitem{mckinney2020international}
S.~M. McKinney, M.~Sieniek, V.~Godbole, J.~Godwin, N.~Antropova, H.~Ashrafian,
  T.~Back, M.~Chesus, G.~S. Corrado, A.~Darzi, {\em et~al.}, ``International
  evaluation of an ai system for breast cancer screening,'' {\em Nature},
  vol.~577, no.~7788, pp.~89--94, 2020.

\bibitem{petersen2019deep}
J.~Petersen, P.~F. J{\"a}ger, F.~Isensee, S.~A. Kohl, U.~Neuberger, W.~Wick,
  J.~Debus, S.~Heiland, M.~Bendszus, P.~Kickingereder, {\em et~al.}, ``Deep
  probabilistic modeling of glioma growth,'' in {\em International Conference
  on Medical Image Computing and Computer-Assisted Intervention}, pp.~806--814,
  Springer, 2019.

\bibitem{petersen2021continuous}
J.~Petersen, F.~Isensee, G.~K{\"o}hler, P.~F. J{\"a}ger, D.~Zimmerer,
  U.~Neuberger, W.~Wick, J.~Debus, S.~Heiland, M.~Bendszus, {\em et~al.},
  ``Continuous-time deep glioma growth models,'' in {\em International
  Conference on Medical Image Computing and Computer-Assisted Intervention},
  pp.~83--92, Springer, 2021.

\bibitem{tiulpin2019kneel}
A.~Tiulpin, I.~Melekhov, and S.~Saarakkala, ``Kneel: knee anatomical landmark
  localization using hourglass networks,'' in {\em Proceedings of the IEEE/CVF
  International Conference on Computer Vision Workshops}, pp.~0--0, 2019.

\bibitem{segal2013multicenter}
N.~A. Segal, M.~C. Nevitt, K.~D. Gross, J.~Hietpas, N.~A. Glass, C.~E. Lewis,
  and J.~C. Torner, ``The multicenter osteoarthritis study (most):
  opportunities for rehabilitation research,'' {\em PM \& R: the journal of
  injury, function, and rehabilitation}, vol.~5, no.~8, 2013.

\bibitem{heusel2017gans}
M.~Heusel, H.~Ramsauer, T.~Unterthiner, B.~Nessler, and S.~Hochreiter, ``Gans
  trained by a two time-scale update rule converge to a local nash
  equilibrium,'' in {\em Advances in Neural Information Processing Systems},
  pp.~6626--6637, 2017.

\bibitem{sajjadi2018assessing}
M.~S. Sajjadi, O.~Bachem, M.~Lucic, O.~Bousquet, and S.~Gelly, ``Assessing
  generative models via precision and recall,'' {\em arXiv preprint
  arXiv:1806.00035}, 2018.

\bibitem{kingma2014adam}
D.~P. Kingma and J.~Ba, ``Adam: A method for stochastic optimization,'' {\em
  arXiv preprint arXiv:1412.6980}, 2014.

\bibitem{paszke2019pytorch}
A.~Paszke, S.~Gross, F.~Massa, A.~Lerer, J.~Bradbury, G.~Chanan, T.~Killeen,
  Z.~Lin, N.~Gimelshein, L.~Antiga, {\em et~al.}, ``Pytorch: An imperative
  style, high-performance deep learning library,'' {\em Advances in neural
  information processing systems}, vol.~32, pp.~8026--8037, 2019.

\bibitem{zhang2018translating}
Z.~Zhang, L.~Yang, and Y.~Zheng, ``Translating and segmenting multimodal
  medical volumes with cycle-and shape-consistency generative adversarial
  network,'' in {\em Proceedings of the IEEE conference on computer vision and
  pattern Recognition}, pp.~9242--9251, 2018.

\bibitem{lipton2017precise}
Z.~C. Lipton and S.~Tripathi, ``Precise recovery of latent vectors from
  generative adversarial networks,'' {\em arXiv preprint arXiv:1702.04782},
  2017.

\bibitem{tiulpin2019multimodal}
A.~Tiulpin, S.~Klein, S.~M. Bierma-Zeinstra, J.~Thevenot, E.~Rahtu, J.~van
  Meurs, E.~H. Oei, and S.~Saarakkala, ``Multimodal machine learning-based knee
  osteoarthritis progression prediction from plain radiographs and clinical
  data,'' {\em Scientific reports}, vol.~9, no.~1, pp.~1--11, 2019.

\bibitem{googlenaturelungcancer}
D.~Ardila, A.~P. Kiraly, S.~Bharadwaj, B.~Choi, J.~J. Reicher, L.~Peng, D.~Tse,
  M.~Etemadi, W.~Ye, G.~Corrado, {\em et~al.}, ``End-to-end lung cancer
  screening with three-dimensional deep learning on low-dose chest computed
  tomography,'' {\em Nature medicine}, vol.~25, no.~6, pp.~954--961, 2019.

\bibitem{kniep2019radiomics}
H.~C. Kniep, F.~Madesta, T.~Schneider, U.~Hanning, M.~H. Sch{\"o}nfeld,
  G.~Sch{\"o}n, J.~Fiehler, T.~Gauer, R.~Werner, and S.~Gellissen, ``Radiomics
  of brain mri: utility in prediction of metastatic tumor type,'' {\em
  Radiology}, vol.~290, no.~2, pp.~479--487, 2019.

\end{thebibliography}

\newpage
\section*{Supplementary Material}
\renewcommand{\thefigure}{S\arabic{figure}}
\renewcommand{\thetable}{S\arabic{table}}
\setcounter{figure}{0}
\setcounter{table}{0}

\begin{table}[h!]
	\centering
	\captionof{table}{\textbf{The knee radiograph datasets for the training, validation, and testing of the predictive model}}
	\scalebox{0.87}{
	\begin{tabular}{c|cccc} \hline
		& \multicolumn{3}{c}{OAI dataset}                                    & MOST dataset  \\ \cline{2-5} 
		                                      & Training \& validation & Test         & \multicolumn{1}{c|}{Total} & External test \\ \hline
		Number of unique patients                   & 3,506                  & 1,290        & 4,796                      & 2,753         \\
		Number of radiographs                       & 44,326                 & 8,655        & 52,981                     & 19,340        \\
		Number of patients without KLS labels        & 852                    & -            & 852                          & -             \\
		Number of radiographs without KLS labels    & 9,849                  & -            & 9,849                          & -             \\ \hline
		Male (\%)                                      & 1,463 (41.7\%)          & 529 (41.0\%) & 1,992 (41.5\%)              & 1,206 (43.8\%) \\
		Age, mean (SD), years                       & 60.6 (9.1)             & 60.3 (9.1)   & 60.7 (9.1)                 & 62.5 (8.1)    \\ \hline
		Patient knees with KLS = 0                  & 2,537                  & 763          & 3,300                      & 2,514         \\
		Patient knees with KLS = 1                  & 1,528                  & 395          & 1,923                      & 1,350         \\
		Patient knees with KLS = 2                  & 2,396                  & 438          & 2,834                      & 1,401         \\
		Patient knees with KLS = 3                  & 1,013                  & 192          & 1,205                      & 1,731         \\
		Patient knees with KLS = 4                  & 321                    & 66           & 387                        & 957           \\
		Knees with OA progression of $\Delta$KLS\textgreater{}1 & 375                    & 95           & 474                        & 661           \\ \hline
	\end{tabular}}
	\caption*{\small
	The semi-quantitative assessments, i.e., Kellgren \& Lawrence scores (KLS) of fixed flexion knee radiographs were available for both OAI and MOST datasets. 
	Abbreviations: SD, standard deviation; KLS, Kellgren \& Lawrence scores.}
	\label{Table: xrays_meta}
\end{table}

\begin{table}[h!]
	\centering
	\captionof{table}{\textbf{Performance metrics of individual radiologists in predicting OA onset and progression as a function of model assistance }}
	\begin{tabular}{ccccccc}
	\hline
	 \begin{tabular}[c]{@{}c@{}}Radiologist \\ (Years of Experience)\end{tabular}& \begin{tabular}[c]{@{}c@{}}Model prediction \\ assistance\end{tabular} & \begin{tabular}[c]{@{}c@{}}Sensitivity\\ (\%)\end{tabular} & \begin{tabular}[c]{@{}c@{}}Specificity\\ (\%)\end{tabular} & \begin{tabular}[c]{@{}c@{}}PPV\\ (\%)\end{tabular} & \begin{tabular}[c]{@{}c@{}}NPV\\ (\%)\end{tabular} & Cohen's Kappa \\ \hline
	\multirow{2}{*}{\begin{tabular}[c]{@{}c@{}}Radiologist 1 \\ (8 years)\end{tabular}} & no & 53.7 & 54.2 & 22.2 & 82.8 & \multirow{2}{*}{0.022} \\
	 & yes & 53.7 & 93.3 & 66.2 & 89.2 &  \\ \hline
	\multirow{2}{*}{\begin{tabular}[c]{@{}c@{}}Radiologist 2 \\ (8 years)\end{tabular}} & no & 30.5 & 80.3 & 27.4 & 82.6 & \multirow{2}{*}{0.068} \\
	 & yes & 49.5 & 95.4 & 72.3 & 88.6 &  \\ \hline
	\multirow{2}{*}{\begin{tabular}[c]{@{}c@{}}Radiologist 3\\ (7 years)\end{tabular}} & no & 31.6 & 84.1 & 32.6 & 83.5 & \multirow{2}{*}{0.070} \\
	 & yes & 33.7 & 95.1 & 62.7 & 85.5 &  \\ \hline
	\multirow{2}{*}{\begin{tabular}[c]{@{}c@{}}Radiologist 4\\ (8 years)\end{tabular}} & no & 22.1 & 89.8 & 34.4 & 82.6 & \multirow{2}{*}{0.127} \\
	 & yes & 47.4 & 92.1 & 59.2 & 87.8 &  \\ \hline
	\multirow{2}{*}{\begin{tabular}[c]{@{}c@{}}Radiologist 5\\ (4 years)\end{tabular}} & no & 36.8 & 78.3 & 29.2 & 83.6 & \multirow{2}{*}{0.053} \\
	 & yes & 49.5 & 91.6 & 58.8 & 88.2 &  \\ \hline
	\multirow{2}{*}{\begin{tabular}[c]{@{}c@{}}Radiologist 6\\ (6 years)\end{tabular}} & no & 54.7 & 58.1 & 24.1 & 84.1 & \multirow{2}{*}{0.223} \\
	 & yes & 70.5 & 66.2 & 33.7 & 90.2 &  \\ \hline
	 \multirow{2}{*}{\begin{tabular}[c]{@{}c@{}}Radiologist 7\\ (17 years)\end{tabular}} & no & 65.3 & 61.1 & 29.0 & 87.9 & \multirow{2}{*}{0.048} \\
	 & yes & 56.8 & 86.4 & 50.5 & 89.2 &  \\ \hline
	\end{tabular}
	\caption*{\small
	Cohen's kappa was calculated as a measure of agreement between two tests from the same reader.
	Abbreviations: PPV: positive predictive value, NPV: negative predictive value.}
	\label{Table: readers}
\end{table}

\begin{table}[h!]
	\centering
	\captionof{table}{\textbf{Diagnostic performance metrics of the proposed and supervised models in predicting OA onset and progression on the OAI test set. p-values relate to inter-model comparisons.}}
	\begin{tabular}{lllllll}
		\hline
		\textbf{Prediction} & \textbf{\begin{tabular}[c]{@{}l@{}}ROC-AUC\\ (95\% CI)\end{tabular}} & \textbf{p-value} & \textbf{\begin{tabular}[c]{@{}l@{}}Sensitivity\\ (95\% CI)\end{tabular}} & \textbf{p-value} & \textbf{\begin{tabular}[c]{@{}l@{}}Specificity\\ (95\% CI)\end{tabular}} & \textbf{p-value} \\ \hline
		\textbf{OA progression} &  &  &  &  &  &  \\ \hline
		our model & \begin{tabular}[c]{@{}l@{}}0.689\\ (0.614, 0.759)\end{tabular} & - & \begin{tabular}[c]{@{}l@{}}0.683\\ (0.567, 0.795)\end{tabular} & - & \begin{tabular}[c]{@{}l@{}}0.612\\ (0.581, 0.643)\end{tabular} & - \\ \hline
		supervised model & \begin{tabular}[c]{@{}l@{}}0.524\\ (0.451, 0.597)\end{tabular} & \textless{}0.001 & \begin{tabular}[c]{@{}l@{}}0.611\\ (0.492, 0.727)\end{tabular} & 0.485 & \begin{tabular}[c]{@{}l@{}}0.448\\ (0.416, 0.479)\end{tabular} & 0.101 \\ \hline
	\end{tabular}
	\caption*{\small
	P-values relate to a pairwise comparison between our model and the supervised model}
\label{Table: ourvssup}
\end{table}

\begin{table}[]
	\centering
	\captionof{table}{\textbf{Diagnostic performance metrics of the proposed and supervised models in predicting OA onset and progression on the MOST test set. p-values relate to inter-model comparisons.}}
	\begin{tabular}{lllllll}
		\hline
		\textbf{Prediction} & \textbf{\begin{tabular}[c]{@{}l@{}}ROC-AUC\\ (95\% CI)\end{tabular}} & \textbf{p-value} & \textbf{\begin{tabular}[c]{@{}l@{}}Sensitivity\\ (95\% CI)\end{tabular}} & \textbf{p-value} & \textbf{\begin{tabular}[c]{@{}l@{}}Specificity\\ (95\% CI)\end{tabular}} & \textbf{p-value} \\ \hline
		\textbf{OA progression} &  &  &  &  &  &  \\ \hline
		our model & \begin{tabular}[c]{@{}l@{}}0.643 \\ (0.596, 0.689)\end{tabular} & - & \begin{tabular}[c]{@{}l@{}}0.614 \\ (0.529, 0.699)\end{tabular} & - & \begin{tabular}[c]{@{}l@{}}0.572 \\ (0.539, 0.605)\end{tabular} & - \\ \hline
		supervised model & \begin{tabular}[c]{@{}l@{}}0.536 \\ (0.481, 0.589)\end{tabular} & \textless{}0.001 & \begin{tabular}[c]{@{}l@{}}0.430 \\ (0.341, 0.519)\end{tabular} & 0.001 & \begin{tabular}[c]{@{}l@{}}0.618 \\ (0.585, 0.650)\end{tabular} & 0.428 \\ \hline
	\end{tabular}
	\caption*{\small
	P-values relate to a pairwise comparison between our model and the supervised model}
	\label{Table: ourvssupmost}
\end{table}

\begin{table}[]
	\centering
	\captionof{table}{\textbf{Diagnostic performance metrics of the proposed model in predicting OA onset and progression on the OAI test set with variable amounts of training data.}}
	\begin{tabular}{lllllll}
		\hline
		\textbf{Prediction} & \textbf{\begin{tabular}[c]{@{}l@{}}ROC-AUC\\ (95\% CI)\end{tabular}} & \textbf{p-value} & \textbf{\begin{tabular}[c]{@{}l@{}}Sensitivity\\ (95\% CI)\end{tabular}} & \textbf{p-value} & \textbf{\begin{tabular}[c]{@{}l@{}}Specificity\\ (95\% CI)\end{tabular}} & \textbf{p-value} \\ \hline
		\textbf{OA progression} &  &  &  &  &  &  \\ \hline
		\begin{tabular}[c]{@{}l@{}}our system:\\ 100\% training data\end{tabular} & \begin{tabular}[c]{@{}l@{}}0.689\\ (0.614, 0.759)\end{tabular} & - & \begin{tabular}[c]{@{}l@{}}0.683\\ (0.567, 0.795)\end{tabular} & - & \begin{tabular}[c]{@{}l@{}}0.612\\ (0.581, 0.643)\end{tabular} & - \\ \hline
		\begin{tabular}[c]{@{}l@{}}our system:\\ 50\% training data\end{tabular} & \begin{tabular}[c]{@{}l@{}}0.657 \\ (0.581, 0.731)\end{tabular} & 0.449 & \begin{tabular}[c]{@{}l@{}}0.611 \\ (0.492, 0.724)\end{tabular} & 0.493 & \begin{tabular}[c]{@{}l@{}}0.655 \\ (0.624, 0.686)\end{tabular} & 0.631 \\ \hline
		\begin{tabular}[c]{@{}l@{}}our system:\\ 10\% training data\end{tabular} & \begin{tabular}[c]{@{}l@{}}0.571 \\ (0.489, 0.652)\end{tabular} & 0.008 & \begin{tabular}[c]{@{}l@{}}0.557 \\ (0.434, 0.675)\end{tabular} & 0.177 & \begin{tabular}[c]{@{}l@{}}0.573 \\ (0.542, 0.605)\end{tabular} & 0.686 \\ \hline
		\begin{tabular}[c]{@{}l@{}}our system:\\ 1\% training data\end{tabular} & \begin{tabular}[c]{@{}l@{}}0.530 \\ (0.460, 0.600)\end{tabular} & \textless{}0.001 & \begin{tabular}[c]{@{}l@{}}0.600 \\ (0.482, 0.717)\end{tabular} & 0.437 & \begin{tabular}[c]{@{}l@{}}0.492 \\ (0.460, 0.524)\end{tabular} & 0.244 \\ \hline
	\end{tabular}
	\caption*{\small
	P-values relate to a pairwise comparison between the model given full data and the models with partial data}
\label{Table: ourd}
\end{table}

\begin{table}[]
	\centering
	\captionof{table}{\textbf{Diagnostic performance metrics of the proposed model in predicting OA onset and progression on the OAI test set with different numbers of nearest neighbors considered.}}
	\begin{tabular}{lllllll}
		\hline
		\textbf{Prediction} & \textbf{\begin{tabular}[c]{@{}l@{}}ROC-AUC\\ (95\% CI)\end{tabular}} & \textbf{p-value} & \textbf{\begin{tabular}[c]{@{}l@{}}Sensitivity\\ (95\% CI)\end{tabular}} & \textbf{p-value} & \textbf{\begin{tabular}[c]{@{}l@{}}Specificity\\ (95\% CI)\end{tabular}} & \textbf{p-value} \\ \hline
		\textbf{OA progression} &  &  &  &  &  &  \\ \hline
		\begin{tabular}[c]{@{}l@{}}our system:\\ \#nn=1\end{tabular} & \begin{tabular}[c]{@{}l@{}}0.689\\ (0.614, 0.759)\end{tabular} & - & \begin{tabular}[c]{@{}l@{}}0.683\\ (0.567, 0.795)\end{tabular} & - & \begin{tabular}[c]{@{}l@{}}0.612\\ (0.581, 0.643)\end{tabular} & - \\ \hline
		\begin{tabular}[c]{@{}l@{}}our system:\\ \#nn=2\end{tabular} & \begin{tabular}[c]{@{}l@{}}0.655 \\ (0.579, 0.730)\end{tabular} & 0.439 & \begin{tabular}[c]{@{}l@{}}0.559 \\ (0.441, 0.678)\end{tabular} & 0.221 & \begin{tabular}[c]{@{}l@{}}0.692 \\ (0.662, 0.721)\end{tabular} & 0.459 \\ \hline
		\begin{tabular}[c]{@{}l@{}}our system:\\ \#nn=5\end{tabular} & \begin{tabular}[c]{@{}l@{}}0.633 \\ (0.558, 0.703)\end{tabular} & 0.196 & \begin{tabular}[c]{@{}l@{}}0.494 \\ (0.371, 0.613)\end{tabular} & 0.064 & \begin{tabular}[c]{@{}l@{}}0.744 \\ (0.715, 0.771)\end{tabular} & 0.203 \\ \hline
		\begin{tabular}[c]{@{}l@{}}our system:\\ \#nn=10\end{tabular} & \begin{tabular}[c]{@{}l@{}}0.645 \\ (0.574, 0.715)\end{tabular} &  0.314 & \begin{tabular}[c]{@{}l@{}}0.643 \\ (0.526, 0.757)\end{tabular} & 0.696 & \begin{tabular}[c]{@{}l@{}}0.595\\ (0.563, 0.626)\end{tabular} & 0.880 \\ \hline
	\end{tabular}
	\caption*{\small
	P-values relate to a pairwise comparison between the model using one NN and the models using multiple NNs}
\label{Table: ournn}
\end{table}

\begin{figure}[h!]
	\centering
	\scalebox{1.0}{
		\includegraphics[trim=100 470 120 0, clip, width=\textwidth]
		{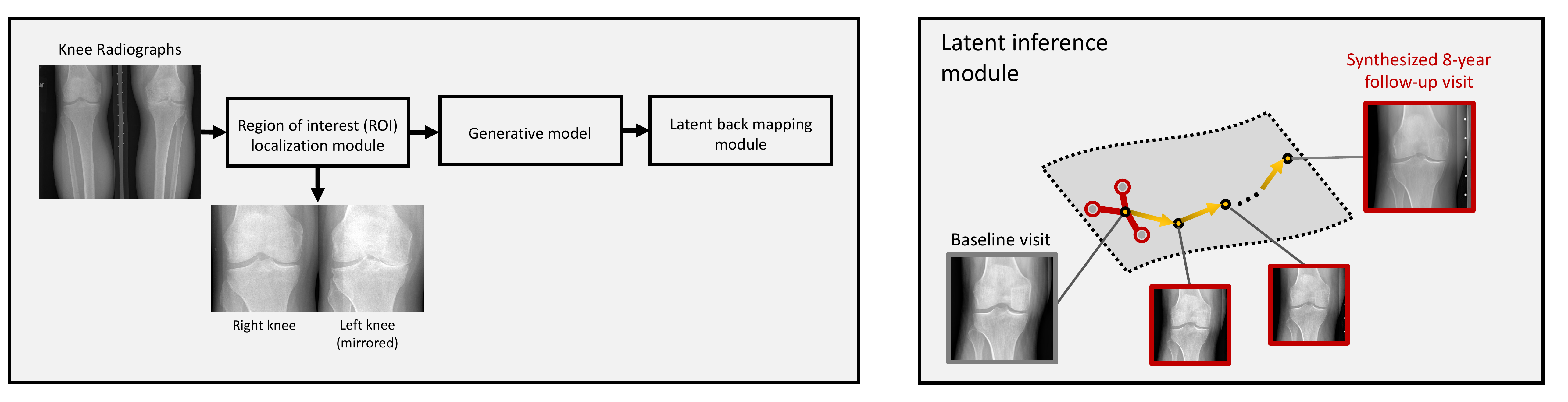}}
	\caption{\label{fig:system}\textbf{ Schematic Overview of Model-Based OA Predictions on Radiographs.}
		On the original knee radiographs (bilateral, posteroanterior projections, and fixed flexion), the image region containing the knee joints is identified using an Hourglass network \cite{tiulpin2019kneel}. 
		Knee joints are cropped accordingly, and a generative model is trained to generate synthetic radiographs from a vector in a low-dimensional latent space. 
		Corresponding original knee X-rays are then back mapped to the point in latent space best matching the synthetic image. 
		In the latent prediction module (right panel), scans of the same patient but at different time points are used to generate a vector field in latent space. 
		A knee radiograph with unknown future OA onset and progression can be mapped to its latent nearest neighbor at a follow-up visit using the vector field. 
	}
\end{figure}

\begin{figure}[h!]
	\centering
	\scalebox{1.0}{
		\includegraphics[trim=0 40 0 50, clip, width=\textwidth]
		{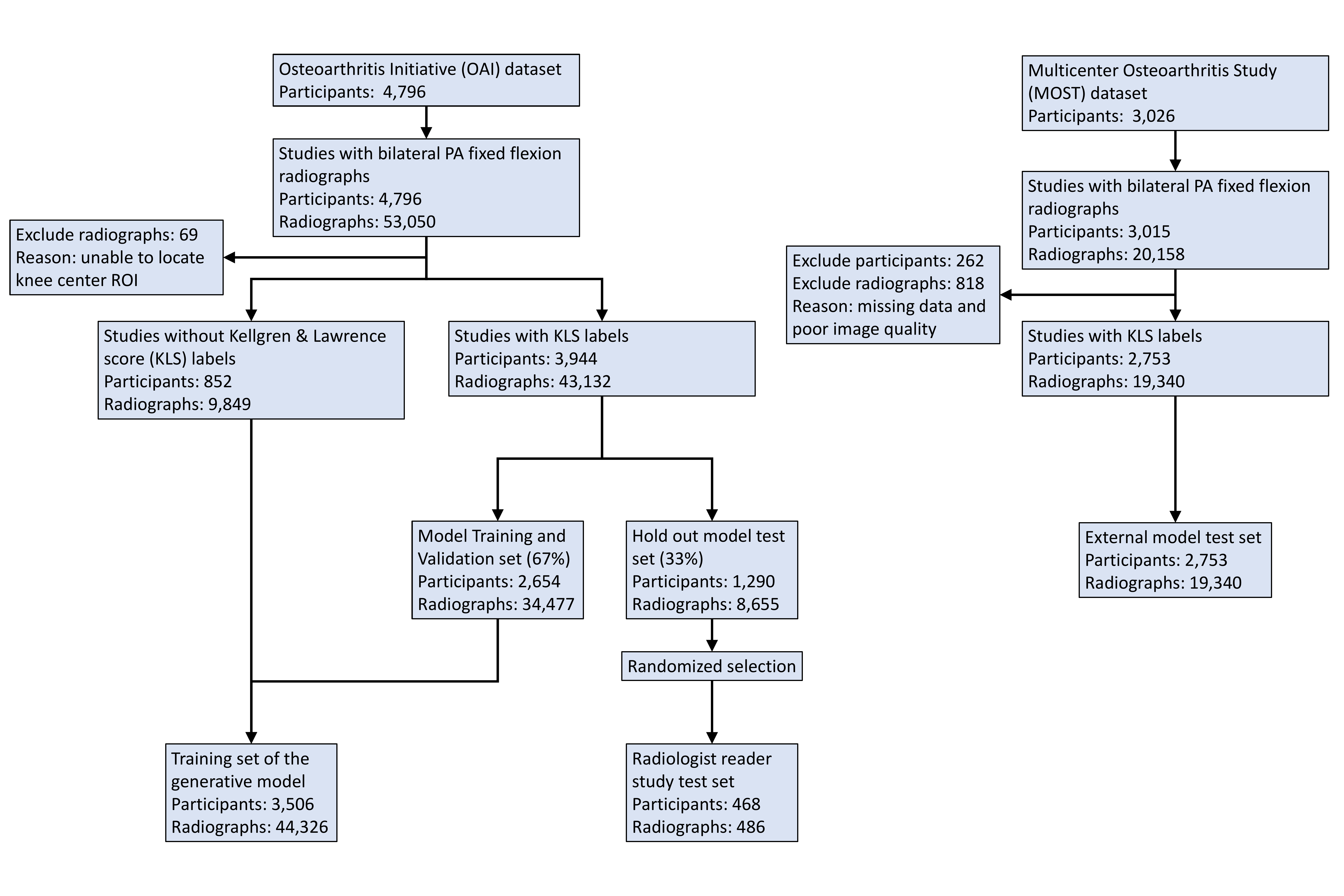}}
	\caption{\label{fig:datapartition}\textbf{Outline of the data processing pipeline.}
		The algorithm was trained on the multi-institutional OAI dataset which was subdivided into a training, validation, and testing set. 
		The hold-out test set was isolated during development and exclusively used for testing. 
		An additional test set was constructed by using the MOST dataset. After excluding knee radiographs with missing information or insufficient image quality, 19,340 radiographs remained on which the algorithm was additionally tested.
	}
\end{figure}

\begin{figure}[h!]
	\centering
	\scalebox{1}{
		\includegraphics[trim=60 100 300 10, clip, width=\textwidth]
		{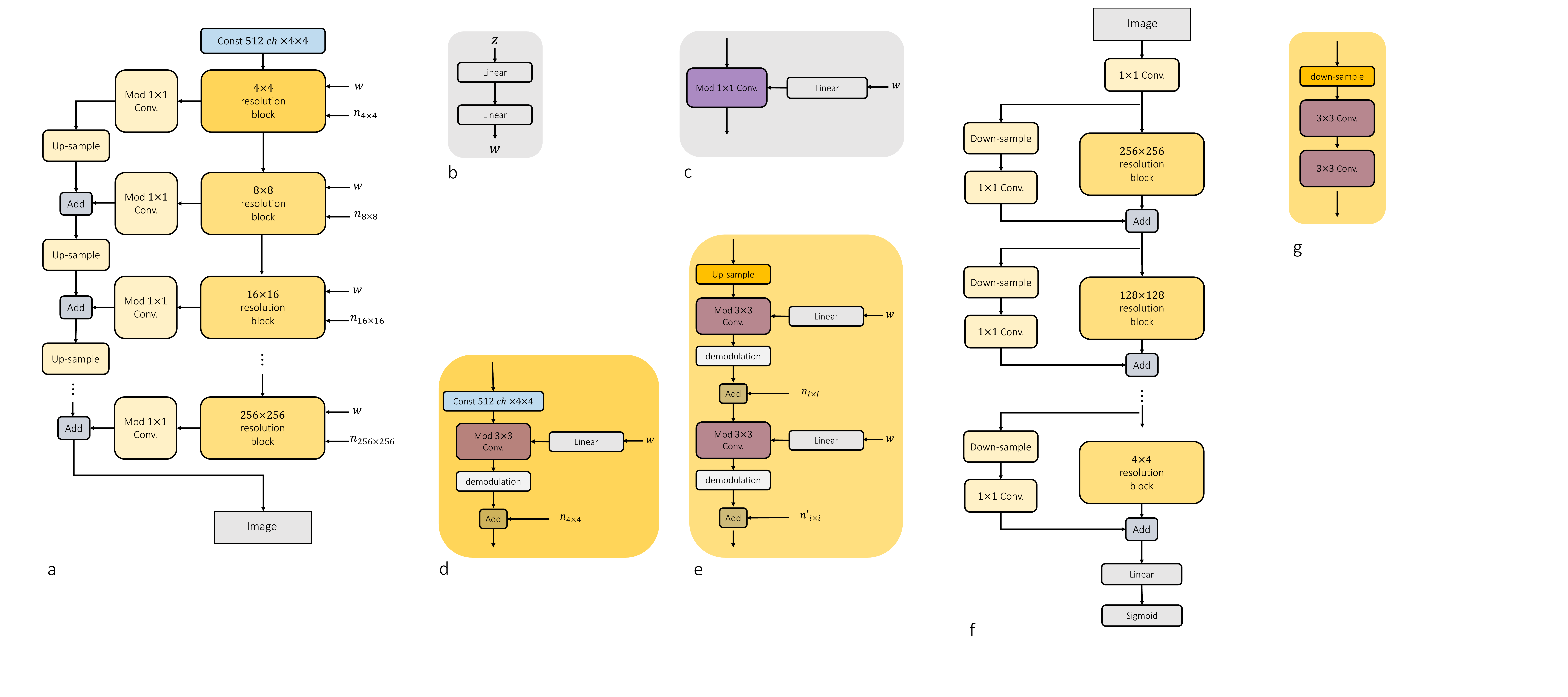}}
	\caption{\label{fig:GAN_nn}\textbf{Architectural details of the generative model.}
	(a) An architectural layout for the StyleGAN's G employed in our work.
	(b) A mapping network consists of 2 hidden linear layers followed by LReLU non-linear activations in G.
	(c) Details of the style modulated 1$\times$1 convolution layer.
	The layer is used to convert feature maps to gray-scale radiographs.
	(d)-(e) Details about resolution blocks.
	The architecture of style-modulated 4$\times$4 resolution block and higher blocks are shown in d and e, respectively.
	(f) An architectural layout for the used StyleGAN's D.
	To increase the performance of D, residual connections are added to fuse features of the previous resolution stage and the current  ones.
	(g) Details of resolution blocks in D.
	}
\end{figure}

\begin{figure}[h!]
	\centering
	\scalebox{0.9}{
		\includegraphics[trim=50 150 310 50, clip, width=\textwidth]
		{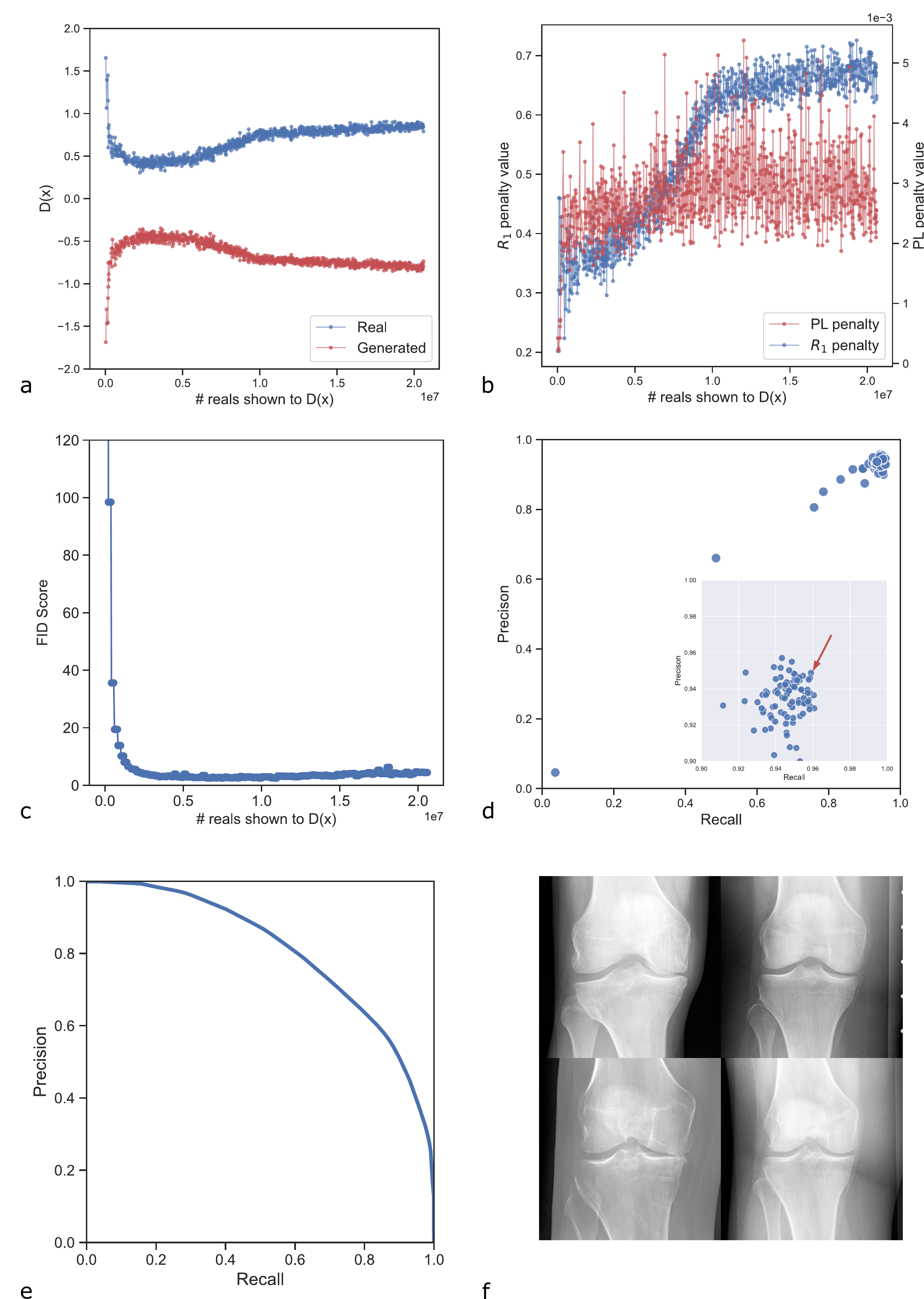}}
	\caption{\label{fig:GAN_train}\textbf{Details of the training process and performance metrics of GAN.}
	 (a) Scores for real (blue) and synthetic radiographs along the training process of the GAN. Scores are stabilized after 10 million images. 
	 (b) Values of G (PL penalty) and D ($R_1$ penalty) penalties along the training process. 
	 (c) Fréchet Inception Distance (FID) scores along the training process to measure convergence to realistic radiographs. 
	 The convergence of the GAN is indicated by the FID scores plateauing after 10 million images are shown to the Discriminator. 
	 (d) Precision and recall metrics for different GAN models. 
	 The model with the highest precision and recall scores (as indicated by the red arrow) was selected. 
	 Inset tiled plot is a close-up of the right upper corner. 
	 (e) Further characterization of the best-performing GAN model via the corresponding distributional precision and recall curve \cite{sajjadi2018assessing}. 
	 (f) Examples of GAN-generated knee radiographs with a spatial resolution of 256$\times$256 pixels.
	}
\end{figure}

\begin{figure}[h!]
	\centering
	\scalebox{1}{
		\includegraphics[trim=60 470 250 0, clip, width=\textwidth]
		{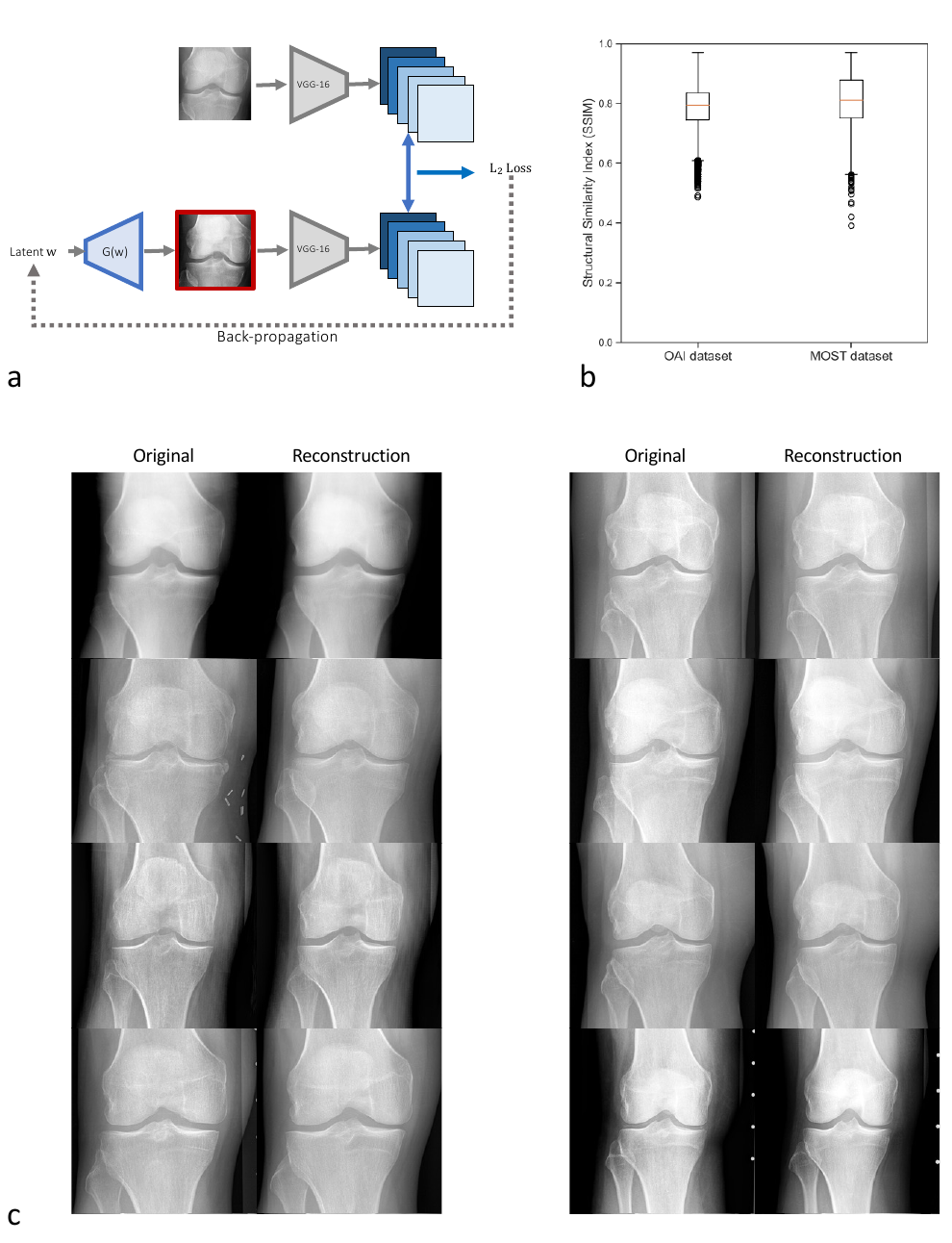}}
	\caption{\label{fig:inverseG}\textbf{Workflow pipeline to visualize the embedding procedure of a representative knee radiograph into the latent space.}
	(a) After computing the learned perceptual image patch similarity (LPIPS) loss, we freeze the parameters in G and the pretrained VGG net to exclusively optimize the intermediate latent $w$ via back propagation. Representative original and synthetic radiographs are framed in gray and red, respectively. 
	(b) In both OAI and MOST datasets, we characterize the embedding procedure by computing the Structural Similarity Index (SSIM) between original and synthetic radiographs. Lines and boxes indicate medians and upper or lower quartile, while whiskers detail the range of the data. 
	(c) Selected reconstructions from the OAI StyleGAN model. 
	As shown, anatomical regions such as the tibia, femur, and knee joint are consistent between original and reconstructed radiographs.
	\todo[inline]{@Daniel, related to R5, we need to add some sentence to the reconstruction evaluation.}
	}
\end{figure}

\begin{figure}[h!]
	\centering
	\scalebox{0.8}{
		\includegraphics[trim=10 160 600 0, clip, width=\textwidth]
		{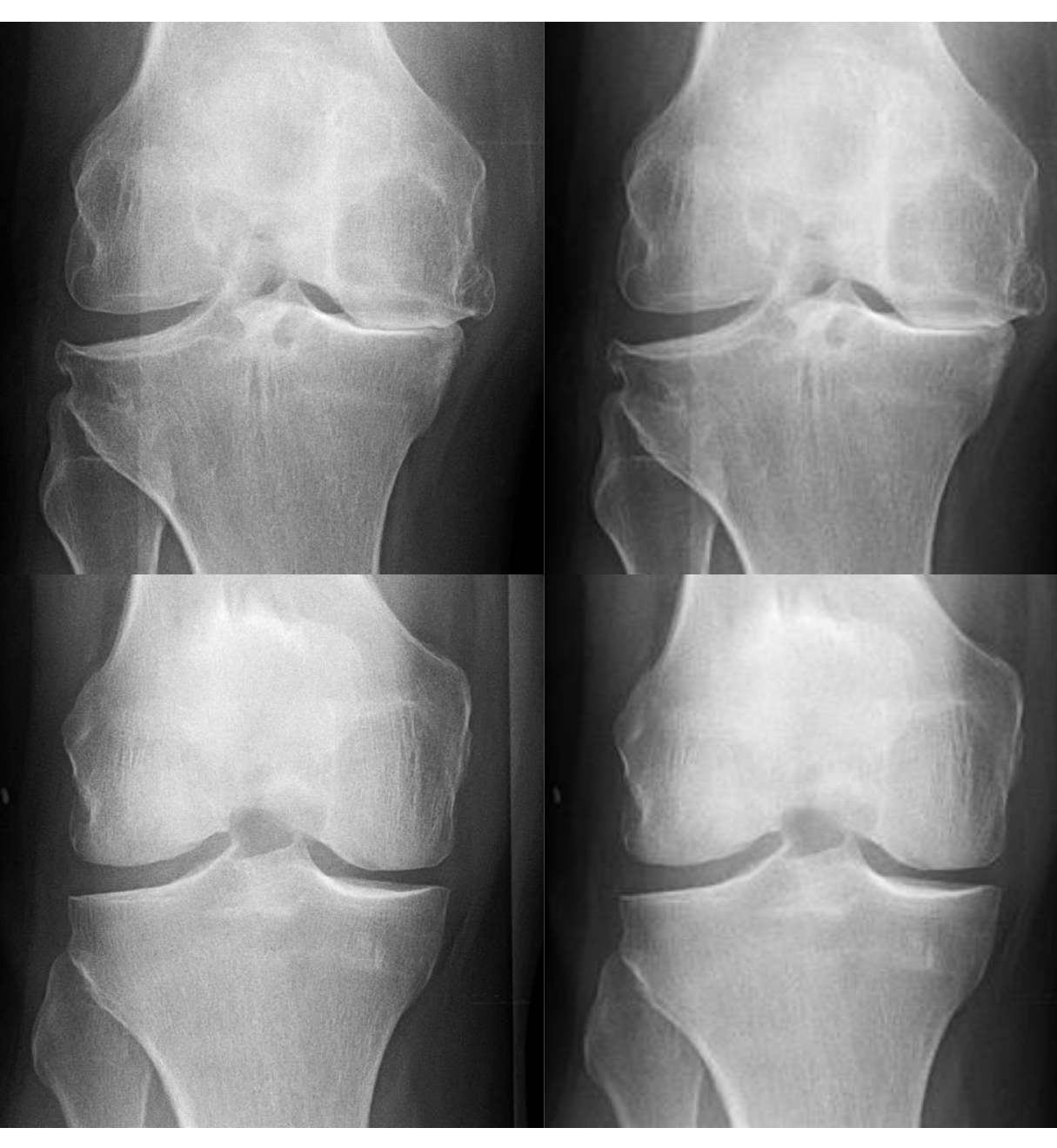}}
	\caption{\label{fig:256vs512}\textbf{Comparison of knee radiographs before and after resizing operation.}
		This example is taken from a 56-year-old male participant. 
		Both left and right knee joints are shown on the top and bottom row, separately. 
		To compare, knee joints with full resolution, i.e., $512\times512$, and with resized $256\times256$ resolution are shown in the left and right columns. 
		In resized radiographs, the diagnostic region around the joint space is still sharp and preserves high image quality. 
		Additional reader studies on full resolution radiographs will be updated soon.      
	}
\end{figure}

\begin{figure}[h!]
	\centering
	\scalebox{1}{
		\includegraphics[trim=0 80 300 0, clip, width=\textwidth]
		{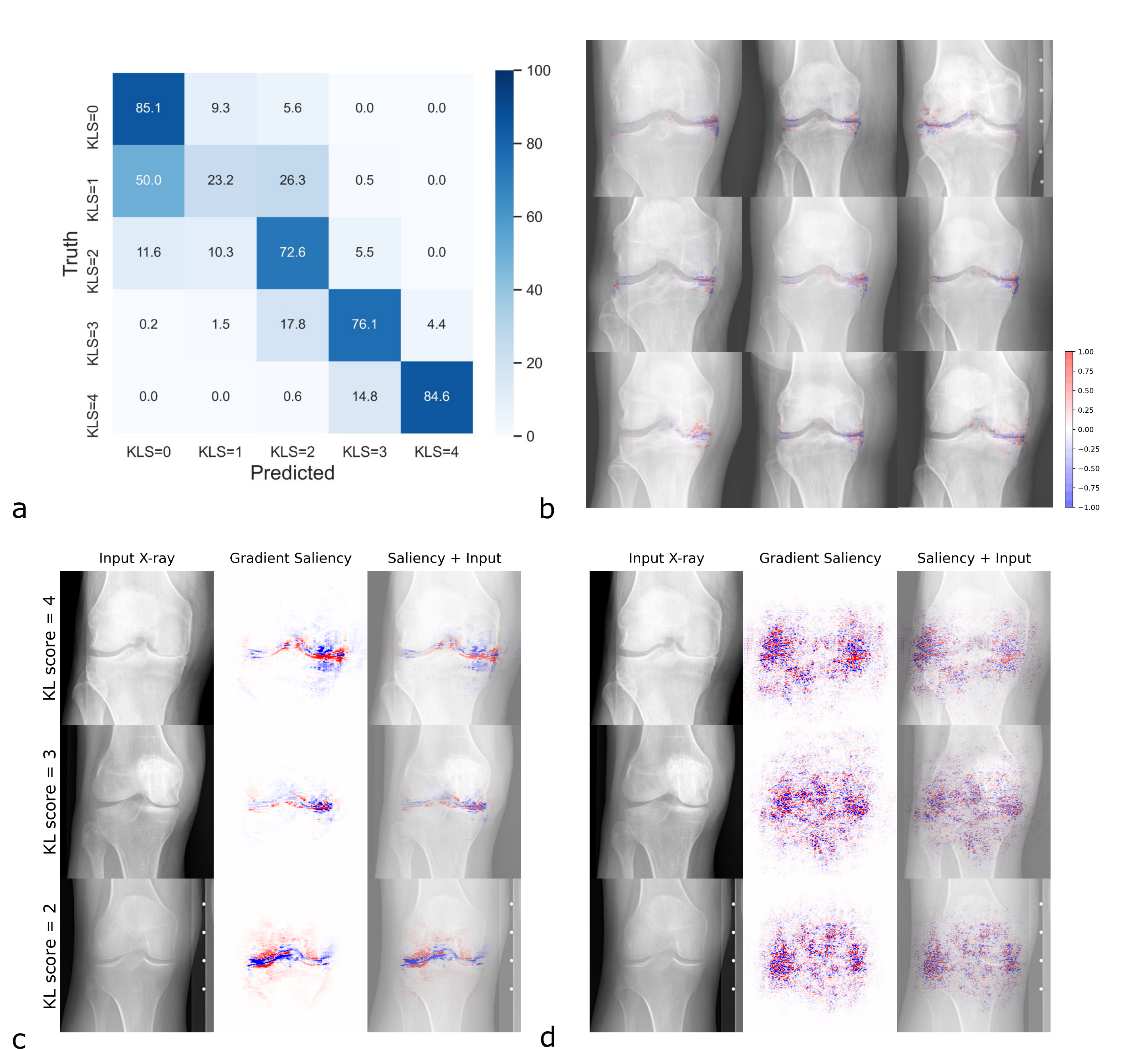}}
	\caption{\label{fig:grads}\textbf{Details about the KLS classifier.}
		(a) Confusion matrix of KLS classification predicted by the ResNet-50 model.  
		(b) Loss gradients were plotted with respect to their input pixels for nine radiographs of the OAI test set displaying signs of OA. 
		Note, the saliency maps created based on the classifier correctly highlight pathology regions.
		(c) and (d), loss gradients with respect to their input pixels for three radiographs displaying signs of OA. 
		Red and blue colored pixels denote positive and negative gradient values.
		To compare, we find the saliency of adversarially trained classifier (in c) aligns well with clinical experts when diagnosing OA thus indicating a superior interpretability \cite{han2021advancing, tsipras2018robustness}, whereas its standard counterpart only generates unrelated and almost noise-like saliences. 
	}
\end{figure}

\begin{figure}[h!]
	\centering
	\scalebox{1.0}{
		\includegraphics[trim=15 490 440 20, clip, width=\textwidth]
		{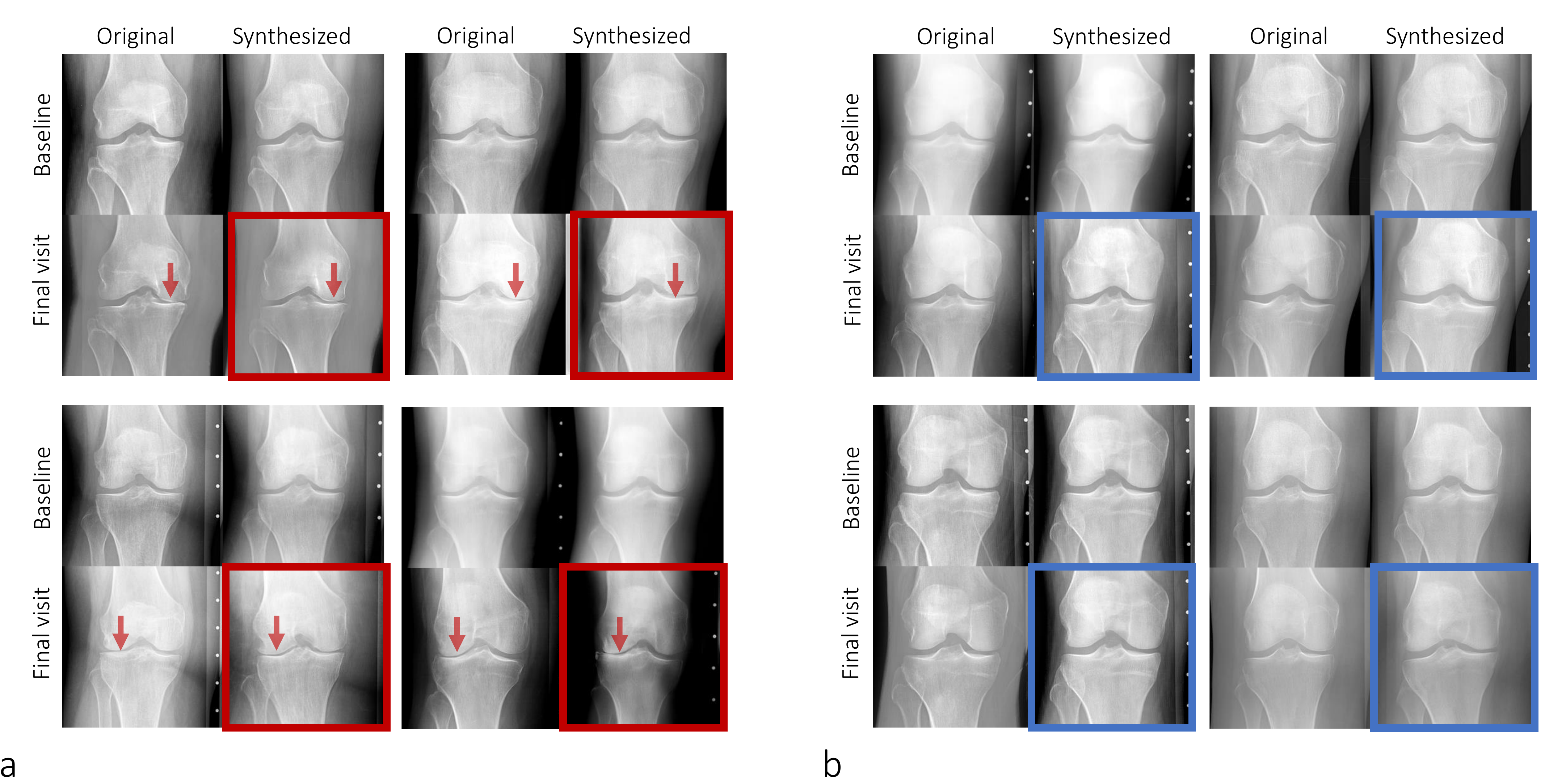}}
	\caption{\label{fig:manifold}\textbf{Model correctly predicts knee joints that exhibit progressive OA development.} 
	Each knee joint is visualized in a 2$\times$2 grid where the ground truth and the model predicted (highlighted in red frames) eight years follow-up outcome are shown in the lower left and lower right panel, respectively.
	As a sign of progressive OA, joint space narrowing, and subchondral opacification are indicated by the red arrows in (a).
	Note that the increasing severity of these OA features is correctly predicted by the model in (a).
	For comparison, in (b), no sign of OA progression is observed in OA non-progressive participants. 
	}
\end{figure}

\begin{figure}[h!]
	\centering
	\scalebox{1.0}{
		\includegraphics[trim=20 20 550 10, clip, width=\textwidth]
		{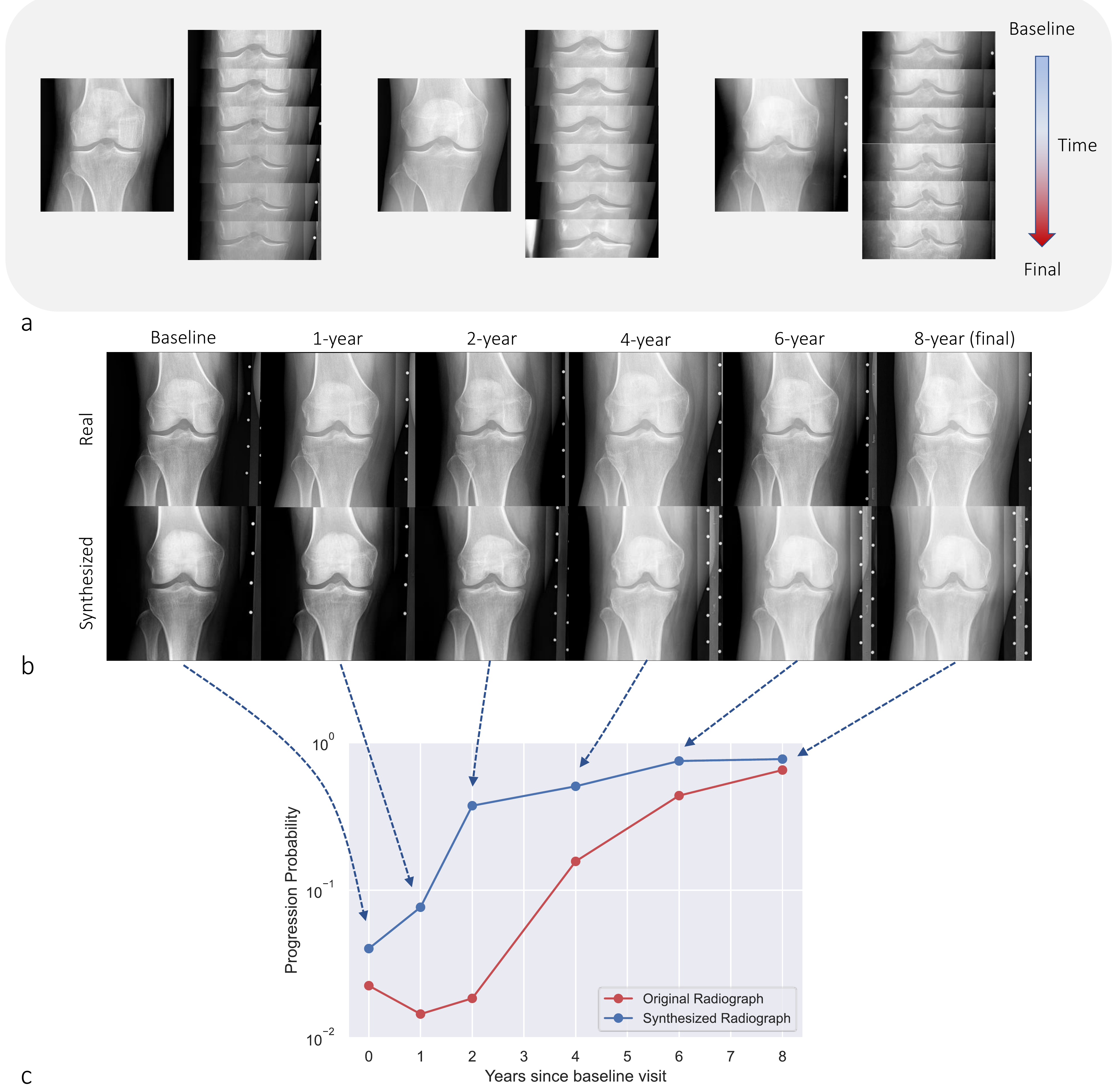}}
	\caption{\label{fig:vis}\textbf{longitudinal prediction of participants with OA progression.}
		(a) Predicted longitudinal evolution of three OA progressive participants from the OAI dataset. 
		(b) Temporal evolution of the left knee of a 61-year-old female participant from the OAI. 
		This was read as displaying no signs of OA at baseline (KLS=0) and presented with mild OA (KLS=2) after 8 years. 
		Original radiographs (upper row) and corresponding model-based synthetic predictions (lower row) at baseline as well as at the 1-year, 2-year, 4-year, 6-year, and 8-year follow-up visits. 
		Original and synthetic radiographs show similar intra-articular changes. 
		Please also note the similar periarticular changes in terms of increased soft tissue around the joint. 
		(c) Based on the images, the respective progression probability was computed for the synthetic (blue) and original radiographs (red). 
		Both curves correctly arrive at a high probability of OA progression, while the synthesized radiographs seem to precede the changes as compared to the original radiographs.
	}
\end{figure}


\begin{figure}[h!]
	\centering
	\scalebox{1.0}{
		\includegraphics[trim=0 140 60 0, clip, width=\textwidth]
		{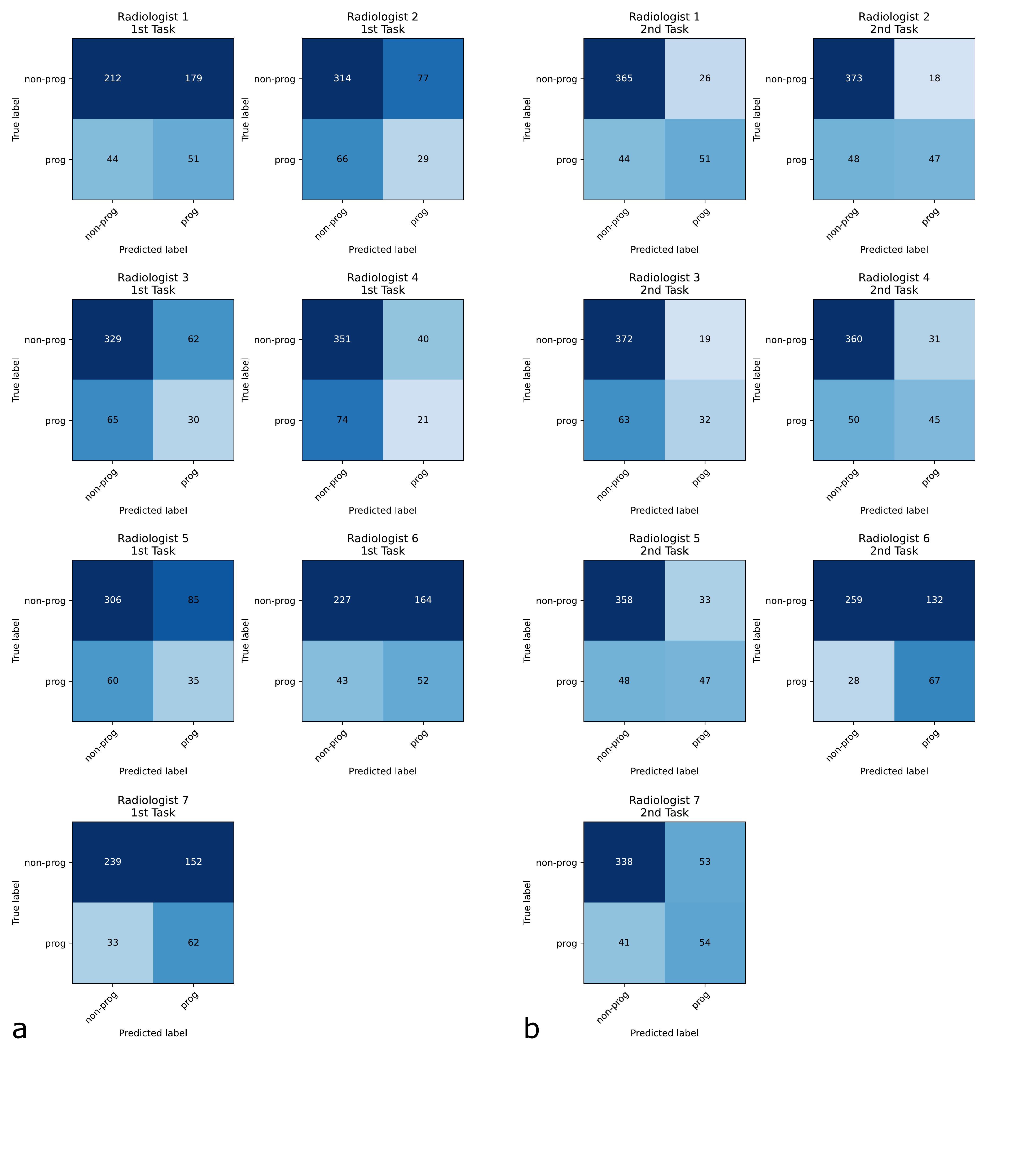}}
	\caption{\label{fig:readerconfusion}\textbf{Confusion matrices for each radiologist and experiment.}
	 Seven radiologists were tasked to judge whether a given knee radiograph is likely to undergo OA progression in the future (prog) or not (non-prog), based on the baseline radiograph alone (a) or on the baseline radiograph and the synthesized predicted radiograph (b). 
	 For both experiments, we randomly selected 486 knee radiographs from the OAI test set.
	 }
\end{figure}

\begin{figure}[h!]
	\centering
	\scalebox{0.8}{
		\includegraphics[trim=200 300 850 200, clip, width=\textwidth]
		{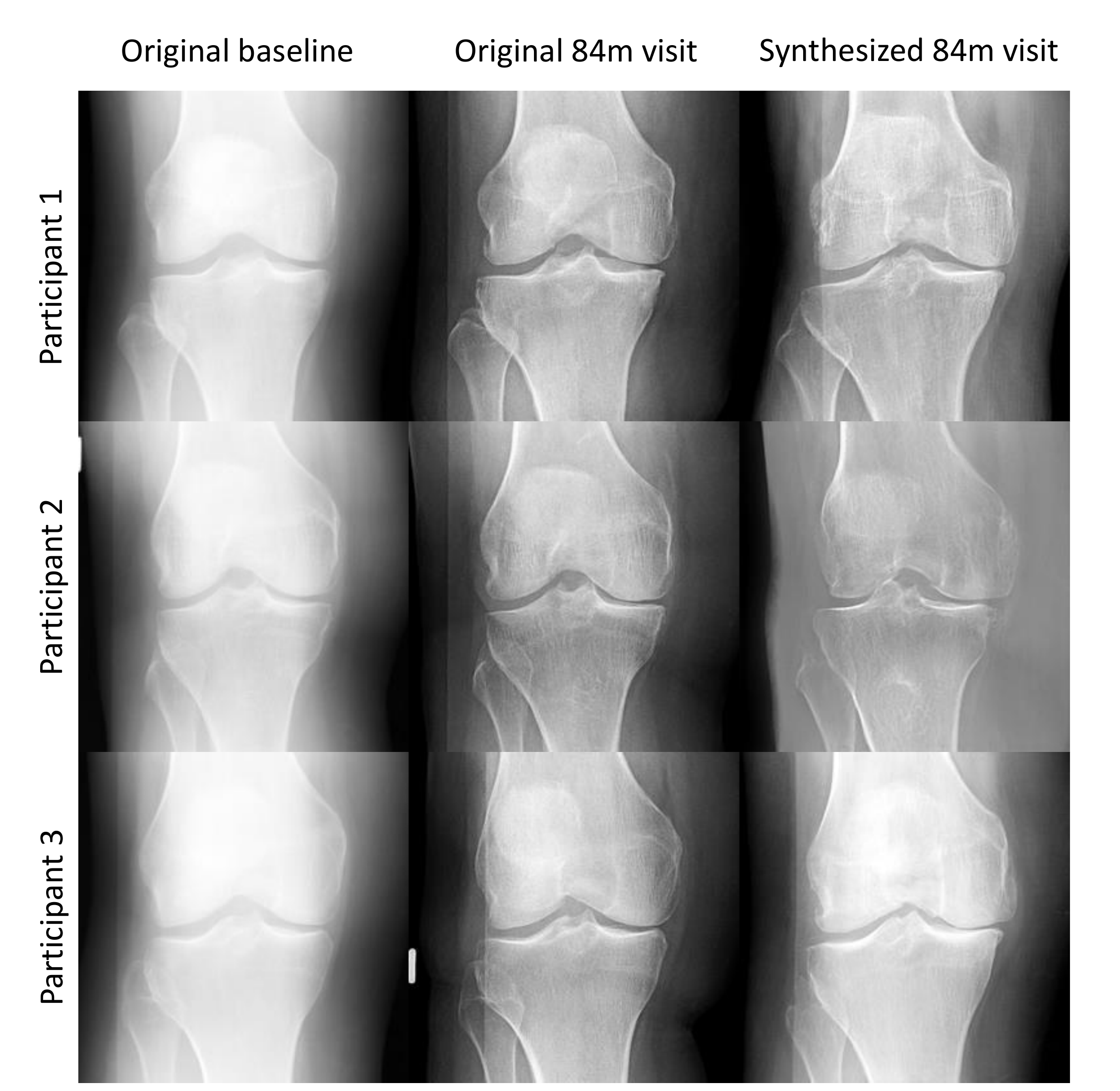}}
	\caption{\label{fig:most}\textbf{Detailed performance metrics of the predictive model on the multi-center osteoarthritis study (MOST) dataset.}
	 Because the predictive model had been developed on data from the OAI, its predictive performance was additionally evaluated on unseen corresponding knee joint radiographs from the MOST dataset containing 2,753 participants and 19,340 radiographs. 
	 Three representative participants with OA progression (listed row-wise). 
	 In the 3$\times$3 grids, the 1st and 2nd columns present the original knee radiographs acquired at baseline and at the 84-month follow-up visit, while the 3rd column presents the predicted (synthetic) future radiograph. 
	 Even though some differences are present, the predicted radiographs demonstrate substantial correspondence with the original follow-up radiographs in terms of a clear OA signature. 
	}
\end{figure} 

\begin{figure}[h!]
	\centering
	\scalebox{1.0}{
		\includegraphics[trim=25 350 20 300, clip, width=\textwidth]
		{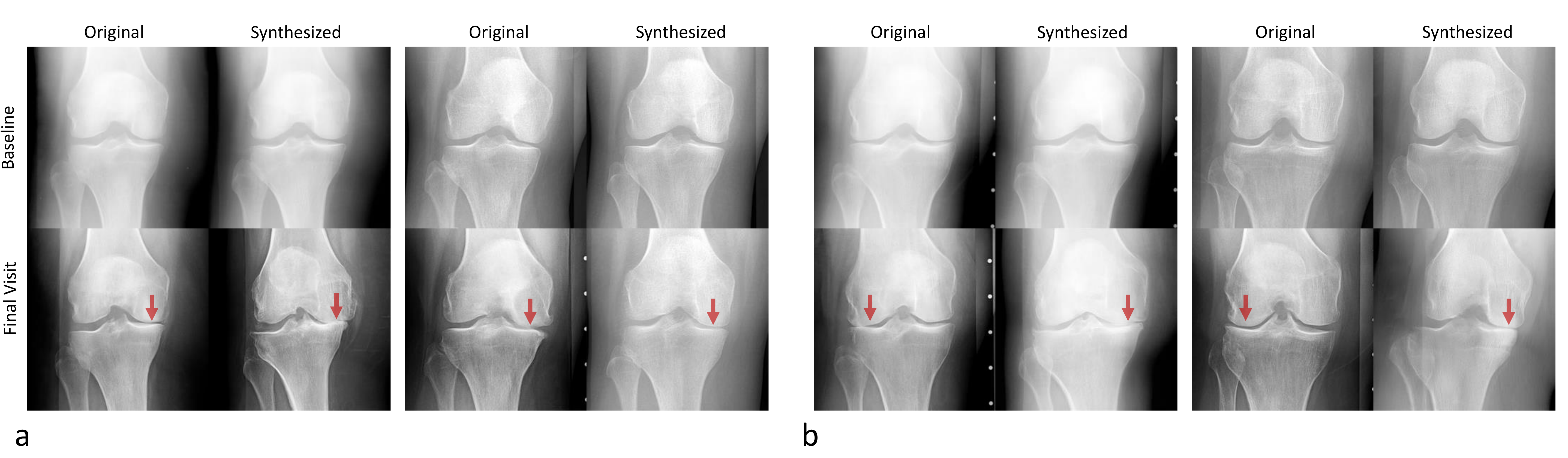}}
	\caption{\label{fig:failure}\textbf{Comparison of correctly and incorrectly predicted future radiographs.}
	Examples of Correctly (a) and Incorrectly (b) Predicted Radiographs. 
	Original radiographs (left column of each 4x4 array) and synthesized radiographs (right column of each 4x4 array).
	While many synthesized radiographs correctly predicted the future course of OA (a), the predictive model failed in some instances and did not correctly predict progression altogether or predicted progression in the wrong compartment, e.g., medial instead of lateral (b). 
	Joint space width narrowing is indicated by arrows. 
	All examples were selected from the OAI test set.
	}
\end{figure} 


\end{document}